\documentclass[a4paper,11pt]{article}
\pdfoutput=1 % if your are submitting a pdflatex (i.e. if you have
\usepackage{jheppub} % for details on the use of the package, please
\usepackage[T1]{fontenc} % if needed
\RequirePackage[displaymath]{lineno} % Display line numbers

\usepackage{epsfig}
\usepackage{graphicx}% Include figure files
\usepackage{dcolumn}% Align table columns on decimal point
\usepackage{bm}% bold math
\usepackage{ltablex,booktabs}
\usepackage{overpic}
\usepackage{subfigure}
\usepackage{float}
\usepackage{color}
\usepackage{amsmath}
\usepackage{mathcomp}
\usepackage{mathrsfs}
\usepackage{multirow}
\usepackage{rotating}
\usepackage{amssymb}
\usepackage{gensymb}
\usepackage{amsmath}
\usepackage{tabularx}
\usepackage{epstopdf}
\newcommand{\phimv}{D_s^+\to \phi\mu^+\nu_\mu}
\newcommand{\phiev}{D_s^+\to \phi e^+\nu_e}
\newcommand{\kkmv}{D_s^+\to K^+K^-\mu^+\nu_\mu}

\title{Studies of the decay  $D^+_s\to K^+K^- \mu^+ \nu_{\mu}$ }
\collaborationImg{\includegraphics[width=0.15\textwidth, angle=90]{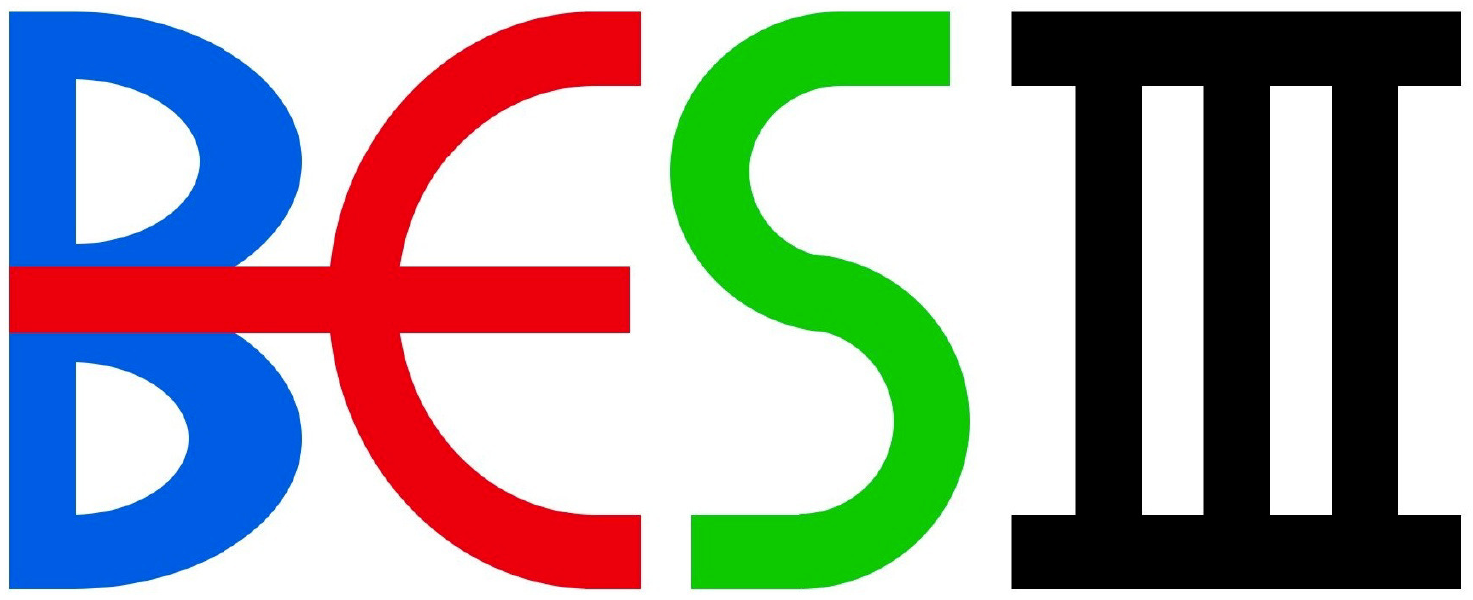}}
\collaboration{The BESIII Collaboration}

\date{\today}
\abstract{The $D^+_s\to K^+K^-\mu^+\nu_\mu$ decay  is studied based on 7.33 fb$^{-1}$ of $e^+e^-$ collision data  collected with the BESIII detector at   center-of-mass energies in the range from 4.128 to 4.226 GeV. The absolute branching fraction is measured as ${\mathcal B}(D^+_s\to \phi \mu^+\nu_\mu) = (2.25\pm 0.09 \pm 0.07) \times10^{-2}$, the most precise measurement to date. Combining with the world average of ${\mathcal B}(D^+_s\to \phi e^+\nu_e)$, the ratio of the branching fractions obtained is$\frac{{\mathcal B}(D^+_s\to \phi \mu^+\nu_\mu)}{{\mathcal B}(D^+_s\to \phi e^+\nu_e)} = 0.94\pm0.08$, in agreement with  lepton universality. By performing a partial wave analysis, the hadronic form factor ratios  at $q^{2}=0$ are extracted, finding $r_{V}=\frac{V(0)}{A_{1}(0)}=1.58\pm0.17\pm0.02$ and $r_{2}=\frac{A_{2}(0)}{A_{1}(0)}=0.71\pm0.14\pm0.02$, where the first uncertainties are statistical and the second are systematic. No significant $S$-wave contribution from $f_0(980)\to K^+K^-$ is found. The upper limit $\mathcal{B}(D_s^+\to f_0(980)\mu^{+}{\nu}_{\mu}) \cdot{\mathcal B}(f_0(980)\to K^+K^-) < 5.45 \times 10^{-4}$ is set at 90\% confidence level.}

\arxivnumber{}
\begin{document}
\maketitle
\flushbottom

%------------------------------------------------------------------------------
\section{Introduction}
 
The study of charm semileptonic (SL) decays provides valuable information about weak and strong interactions in mesons composed of heavy quarks. The SL partial decay width is related to the product of the hadronic form factors (FFs) describing the strong interactions between final-state quarks, including non-perturbative effects, and the Cabibbo-Kobayashi-Maskawa (CKM) matrix elements. Precise measurements of FFs are important for comparison with the theoretical calculations. From the theoretical point of view, hadronic FFs play a crucial role in the study of  leptonic $D_s$  decays. They are calculated by non-perturbative methods, including lattice quantum chromodynamics (LQCD)~\cite{Na:2011mc,Aoki:2016frl,Donald:2013pea,Donald:2013kla} and phenomenological quark models. The latter category includes the covariant confined quark model (CCQM)~\cite{Soni:2018adu},  the constituent quark model (CQM)~\cite{Melikhov:2000yu}, the light-front quark model (LFQM)~\cite{Verma:2011yw}, as well as the HM$\chi$T model~\cite{Fajfer:2005ug} (based on the combination of heavy meson and chiral symmetries). The $D^+_s\to\phi\ell^+\nu_\ell$ decay is particularly interesting since the $\phi$ meson is a narrow resonance, which can be isolated, providing a good testbed.  

In the Standard Model (SM), SL decays  offer an excellent opportunity to test  lepton flavor universality (LFU) and search for new physics effects. Previously, LHCb tests of LFU using $B ^+ \to K ^ {+} \ell ^+ \ell ^ - $ decays hinted at LFU violation, with a significance of $3.1\sigma$~\cite{LHCb:2021trn}.  
However, more recently LHCb tests of LFU in $B^+ \to K^{+}\ell^+\ell^-$ and $B^0 \to K^{*0}\ell^+\ell^-$ decays~\cite{LHCb:2022zom} are consistent with the SM at the $1.0\sigma$  level. The possible tension is addressed by various theoretical models~\cite{Bauer:2015knc,Crivellin:2015hha,Crivellin:2015mga,Fajfer:2012jt,Fajfer:2012vx}. 
Searches for LFU violation have also been performed in SL decays of $D^{0(+)}$ and $D^+_s$ mesons~\cite{BESIII:2018nzb,BESIII:2019gsm,Ablikim:2020hsc,BESIII:2019qci,BESIII:2017ikf} and baryons~\cite{BESIII:2021ynj,Belle:2021crz,Belle:2021dgc,BESIII:2015ysy}, without any clear evidence for deviation with respect to the SM predictions. Hence, higher precision measurements are desirable. For $D^+_s\to\phi\ell^+\nu_\ell$ decays, $BABAR$  performed the most precise  measurement of the absolute branching fraction (BF) of $D^+_s\to\phi e^+\nu_e$ with an uncertainty of 6.6\%~\cite{BaBar:2008gpr}. In comparison, the uncertainty of the BF of $D^+_s\to\phi \mu^+\nu_\mu$ measured by the BESIII experiment previously is 26.3\%~\cite{BESIII:2017ikf}, which limits the precision of LFU studies using $D^+_s\to\phi\ell^+\nu_\ell$ decays. Therefore, a precision measurement of the absolute BF of $D^+_s\to\phi \mu^+\nu_\mu$ can provide a critical, complementary test for LFU. 

Using the 7.33 fb$^{-1}$ data sample collected by BESIII at center-of-mass energies ($E_{\rm CM}$) in the range from 4.128 to 4.226 GeV, a measurement of the BF of the $\phimv$ decay with significantly improved precision is reported. 
LFU is tested using the world average value of the $\phiev$ BF. 
Additionally, the hadronic FFs of $D^+_s\to\phi \mu^+\nu_\mu$  are extracted through a partial wave analysis (PWA) and the size of a possible $f_0(980)$ component in the decay $\kkmv$ is limited. An $f_0(980)$ contribution would be interesting, in view of the unconventional nature of this state~\cite{Kim:2018zob,Wang:2022vga,Stone:2013eaa,Yao:2020bxx,Dai:2014zta}. 
Charge conjugation is implied throughout this work.

\section{Detector and Monte Carlo simulations}
\label{sec:detector_dataset}
The BESIII detector~\cite{BESIII:2009fln} records symmetric $e^+e^-$ collisions
provided by the BEPCII storage ring~\cite{Yu:2016cof} in the  $E_{\rm CM}$ range from 2.0 to 4.95~GeV, with a peak luminosity of $1 \times 10^{33}\;\text{cm}^{-2}\text{s}^{-1}$ achieved at $E_{\rm CM} = 3.77\;\text{GeV}$. BESIII has collected large data samples in this energy region~\cite{BESIII:2020nme, Li:2021iwf}. The cylindrical core of the BESIII detector covers 93\% of the full solid angle and consists of a helium-based multilayer drift chamber~(MDC), a plastic scintillator time-of-flight system~(TOF), and a CsI(Tl) electromagnetic calorimeter~(EMC), which are all enclosed in a superconducting solenoidal magnet providing a 1.0~T magnetic field~\cite{Huang:2022wuo}. The solenoid is supported by an octagonal flux-return yoke with resistive plate counter muon identification modules interleaved with steel. The charged-particle momentum resolution at $1~{\rm GeV}/c$ is $0.5\%$, and the resolution of the specific ionization energy loss  ($\mathrm{d}E/\mathrm{d}x$)  is $6\%$ for electrons from Bhabha scattering. The EMC measures photon energies with a resolution of $2.5\%$ ($5\%$) at $1$~GeV in the barrel (end cap) region. The time resolution in the TOF barrel region is 68~ps. The end cap TOF system was upgraded in 2015 using multi-gap resistive plate chamber technology, providing a time resolution of 60~ps~\cite{Cao:2020ibk}.  
About 83\% of the dataset used in this analysis benefits from this upgrade.  

Simulated data samples produced with a {\sc geant4}-based~\cite{GEANT4:2002zbu} Monte Carlo (MC) package, which includes the geometric description of the BESIII detector and the detector response, are used to determine detection efficiencies and to estimate backgrounds. The simulation models the beam energy spread and initial state radiation (ISR) in the $e^+e^-$ annihilations with the generator {\sc kkmc}~\cite{Jadach:2000ir}.  An  inclusive MC sample with a luminosity equivalent to 40 times that of data is generated at $E_{\rm CM}\in [4.128,4.226]$~GeV.  This MC is used to determine the distributions of kinematic variables and estimate the detection efficiency.
It includes the production of open charm processes, the ISR production of vector charmonium(-like) states, and the continuum processes incorporated in {\sc kkmc}~\cite{Jadach:2000ir}. The production of open charm states directly via $e^+e^-$ annihilations is modeled with the generator {\sc conexc}~\cite{Ping:2013jka}, and their subsequent decays are modeled by {\sc evtgen}~\cite{Lange:2001uf, Ping:2008zz} with known BFs from the Particle Data Group (PDG)~\cite{ParticleDataGroup:2020ssz}. The ISR production of vector charmonium(-like) states and the continuum processes are incorporated in {\sc kkmc}~\cite{Jadach:2000ir}. The remaining unknown charmonium decays are modelled with {\sc lundcharm}~\cite{Chen:2000tv, Yang:2014vra}. Final state radiation (FSR) from charged final-state particles is incorporated using the {\sc photos} package~\cite{Richter-Was:1992hxq}. A phase-space (PHSP) MC sample is produced for  $\kkmv$ and is used to extract the detection efficiency.  Initially, this PHSP MC sample is used to calculate the normalization integral used in the determination of the amplitude model parameters in the fit to data. Then, the signal MC sample is regenerated  with the $D_s^+$ meson decaying to $K^+K^-\mu^+\nu_\mu$ using the fitted amplitude model. It is used to find the final PWA solution and obtain the signal efficiency.

\section{Analysis method}
\label{Double-method}
A double-tag (DT) method  is used in this analysis following Refs.~\cite{BESIII:2021anh,MARK-III:1985hbd,BESIII:2018hhz}.  At $E_{\rm CM}$ between 4.128 and 4.226~GeV,  $D_s$ mesons are mainly produced via the process $e^+ e^- \to D_s^{*+} [{} \to \gamma (\pi^0) D_s^+] D_s^-$. One $D_s^-$ meson is fully reconstructed in  one of the hadronic decay modes, called a single-tag (ST) candidate.  Based on this,  among the particles recoiling against the ST $D_s^-$  meson, we select  the signal decay of the $D_s^+$ meson and a transition $\gamma(\pi^0)$ from the $D_s^{*+}$; success results in a double-tag (DT) candidate. 

To measure the BF of the signal decay, the following equations for one ST mode are used:
\begin{eqnarray}\begin{aligned}
  N_{\text{tag}}^{\text{ST}} = 2N_{D_s^{*\pm}D_{s}^{\mp}}\mathcal{B}_{\text{tag}}\epsilon_{\text{tag}}^{\text{ST}}\,, \label{eq-ST}
\end{aligned}\end{eqnarray}
\begin{equation}
  N_{\text{tag,sig}}^{\text{DT}}=2N_{D_s^{*\pm}D_{s}^{\mp}}\mathcal{B}_{\text{tag}}\mathcal{B}_{\text{sig}}\epsilon_{\text{tag,sig}}^{\text{DT}}\,,
  \label{eq-DT}
\end{equation}
where $N_{\text{tag}}^{\text{ST}}$ is the ST yield for the tag mode, $N_{\text{tag,sig}}^{\text{DT}}$ is the DT yield, 
$N_{D_s^{*\pm}D_{s}^{\mp}}$ is the total number of $D_{s}^{*\pm}D_{s}^{\mp}$ pairs produced in the $e^{+}e^{-}$ collisions, 
$\mathcal{B}_{\text{tag}}$ and $\mathcal{B}_{\text{sig}}$ are the BFs of the tag and signal modes, respectively, $\epsilon_{\text{tag}}^{\text{ST}}$ is the ST efficiency to reconstruct the tag mode and $\epsilon_{\text{tag,sig}}^{\text{DT}}$ is the DT efficiency to reconstruct both the tag and signal decay modes. 
In the case of more than one tag mode and energy point, Eq.~(\ref{eq-DT}) can be written as
\begin{eqnarray}
\begin{aligned}
  \begin{array}{lr}
    N_{\text{total}}^{\text{DT}}=\sum\limits_{\alpha, j}N_{\alpha,\text{sig},j}^{\text{DT}}   = \mathcal{B}_{\text{sig}}
		\sum\limits_{\alpha, j}2N^{j}_{D_s^{*\pm}D_{s}^{\mp}}\mathcal{B}_{\alpha}\epsilon_{\alpha,\text{sig}, j}^{\text{DT}}\,,
  \end{array}
  \label{eq-DTtotal}
\end{aligned}
\end{eqnarray}
where $\alpha$  represents the tag-mode and $j$ is the energy point (from 0 to 7, corresponding to the energy points in Table~\ref{tab:eff2}). 
$\mathcal{B}_{\text{sig}}$ is isolated by using Eq.~(\ref{eq-ST}):
\begin{eqnarray}\begin{aligned}
  \mathcal{B}_{\text{sig}} =
  \frac{N_{\text{total}}^{\text{DT}}}{ \begin{matrix}\sum\limits_{\alpha, j} N_{\alpha, j}^{\text{ST}}\epsilon^{\text{DT}}_{\alpha,\text{sig},j}/\epsilon_{\alpha,j}^{\text{ST}}\cdot\mathcal B_{\rm sub}\end{matrix}},
  \label{eq1}
\end{aligned}\end{eqnarray}
where   $N_{\text{total}}^{\text{DT}}$ denotes the total number of DT events obtained from the fit to the signal peaks (see below) of
the selected DT candidates, while $N_{\alpha,j}^{\text{ST}}$ and $\epsilon_{\alpha,j}^{\text{ST}}$ are obtained from the data and inclusive MC samples, respectively. Finally, $\epsilon_{\alpha,\text{sig},j}^{\text{DT}}$ is determined with signal MC samples. These efficiencies do not include the product of the BFs, $\mathcal B_{\rm sub}$, for the intermediate resonance decays.

\section{Single tag selection}
\label{ST-selection}
Candidates for the  ST $D^-_s$ mesons  are reconstructed via fourteen  hadronic decay modes 
$D^-_s\to K^+K^-\pi^-$, $K^-\pi^+\pi^-$, $\pi^+\pi^-\pi^-$,  $K^+K^-\pi^-\pi^0$, $\eta^\prime_{\gamma\rho^0}\pi^-$,
$\eta_{\gamma\gamma}\rho^-$, $K^0_SK^-\pi^+\pi^-$, $K^0_SK^+\pi^-\pi^-$, $\eta_{\gamma\gamma}\pi^-$,
$K^0_SK^0_S\pi^-$, $\eta_{\pi^0\pi^+\pi^-}\pi^-$, $\eta^\prime_{\eta_{\gamma\gamma}\pi^+\pi^-}\pi^-$,  
$K^0_SK^-\pi^0$, and $K^0_SK^-$. Throughout this paper,  the subscripts of $\eta$ and $\eta^{\prime}$ indicate the decay modes used for reconstructing the corresponding particle. 

The selection criteria of $\pi^\pm$, $K^\pm$, $K^0_S$, $\gamma$, $\pi^0$, and $\eta$ candidates follow Refs.~\cite{BESIII:2019qci,BESIII:2019pjk}. All charged tracks are required to be within  $|\!\cos\theta|<0.93$, where $\theta$ is the polar angle  defined with respect to the symmetry axis of the MDC. For the charged tracks that are not  from $K_S^0$ decays, the distance of closest approach to the interaction point must  be less than 10 cm along the beam direction and less than 1 cm in the plane perpendicular to the beam. Particle identification (PID) of the charged particles is performed  by combining $\mathrm{d}E/\mathrm{d}x$ measurements in the MDC with flight time measurements in the TOF system.
Tracks are identified by the PID likelihood ${\cal L}_{h}$ ($h = \pi, K$) for each hadron $h$ hypothesis. 
 Pion and  kaon candidates are required to satisfy  ${\cal L}_{\pi}>{\cal L}_{K}$ and ${\cal L}_{K}>{\cal L}_{\pi}$, respectively.

The $K_S^0$ candidates are selected by looping over all pairs of tracks with opposite charges, which distances to the interaction point along the beam direction are within 20 cm; these tracks are treated as pions without applying PID. The $\pi^+\pi^-$ invariant mass is required to be in the range of $(0.487,0.511)$ GeV/$c^2$. 
The signed decay length, $L$, of the reconstructed $K_S^0$ is required to be separated from the interaction point by greater than twice its resolution, $\sigma$: $L/\sigma > 2$.  

Photon shower candidates are selected from energy clusters in the EMC that are not associated with any charged track. 
To reduce the number of photon candidates that result from noise and beam background, each shower is required to start within 700\,ns of the event start time. The deposited energy of showers in the barrel region and in the end-cap region must be greater than 25 MeV and 50 MeV~\cite{Ablikim:2009aa}, respectively. To exclude showers that originate from charged tracks, the angle subtended by the EMC shower and the position of the closest charged track at the EMC must be greater than 10 degrees as measured from the interaction point.

%The $\pi^0$ and $\eta$ mesons are reconstructed from photon pairs. 
Photon pairs are used to  reconstruct the $\pi^0$ and $\eta$ mesons.
The invariant masses of the selected photon pairs are required to be
within the intervals $(0.115,\,0.150)$ and $(0.500,\,0.570)$\,GeV$/c^{2}$, respectively. To improve the momentum resolution and suppress background contributions,  a kinematic fit is applied to each selected photon pair, whose  invariant mass is  constrained to the nominal mass of $\pi^{0}$ or $\eta$~\cite{ParticleDataGroup:2020ssz}.

For the  $D^-_s\to \eta_{\pi^0\pi^+\pi^-}\pi^-$  tag mode,  the invariant mass $M_{\pi^0\pi^+\pi^-}$ of the $\pi^0\pi^+\pi^-$ combinations used to form $\eta$ candidates is required to be within the  interval $(0.530,\,0.570)~\mathrm{GeV}/c^2$. 
The  two decay modes $\eta\pi^+\pi^-$ and $\gamma\rho^0$ are used to  reconstruct $\eta^\prime$ candidates, while their invariant masses are required to fall in the ranges of $(0.946,\,0.970)$ GeV/$c^2$ and $(0.940,\,0.976)~\mathrm{GeV}/c^2$, respectively. Additionally, the  energy of the $\gamma$ from $\eta'\to\gamma\rho^0$ decays are required to  be greater than 0.1\,GeV. The $\pi^+\pi^-$ and $\pi^-\pi^0$ combinations are used to form $\rho^0$ and $\rho^-$ candidates, respectively, and their invariant masses are required to fall in the range of $(0.570,\,0.970)~\mathrm{GeV}/c^2$.

To suppress the transition  pions from $D^{*+}\to D^0\pi^+$, the minimum  momenta of all the pions, which are not from the  $K_S^0$, $\eta$, or $\eta^\prime$ decays, must  be greater than 0.1\,GeV/$c$. For the $D^-_s\to K^-\pi^+\pi^-$ and $D^-_s\to \pi^+\pi^-\pi^-$  tag modes, the peaking  background events from $D^-_s\to K^0_SK^-$ and $D^-_s\to K^0_S\pi^-$  are suppressed  by requiring the    $\pi^+\pi^-$ invariant mass  to be at least $0.03$ GeV/$c^2$ away from the known  $K^0_S$ mass~\cite{ParticleDataGroup:2020ssz}.

To reject the  non-$D_s^{\pm}D^{*\mp}_s$ backgrounds,
we define the  beam-constrained mass of the ST $D_s^-$
candidate as 
\begin{equation}
M_{\rm BC}\equiv\sqrt{E_{\rm CM}^2/4-|\vec{p}_{D^-_s}|^2}
  \label{Mbc}
\end{equation}
and require $M_{\rm BC}$ to   be within the region listed in Table~\ref{tab:eff2}. This selection criterion accepts  most of the $D_s^-$ mesons from the  $e^+ e^- \to D_s^{\pm}D_s^{*\mp}$ process. 

\begin{figure*}[htbp]
\centering
\includegraphics[width=0.98\textwidth]{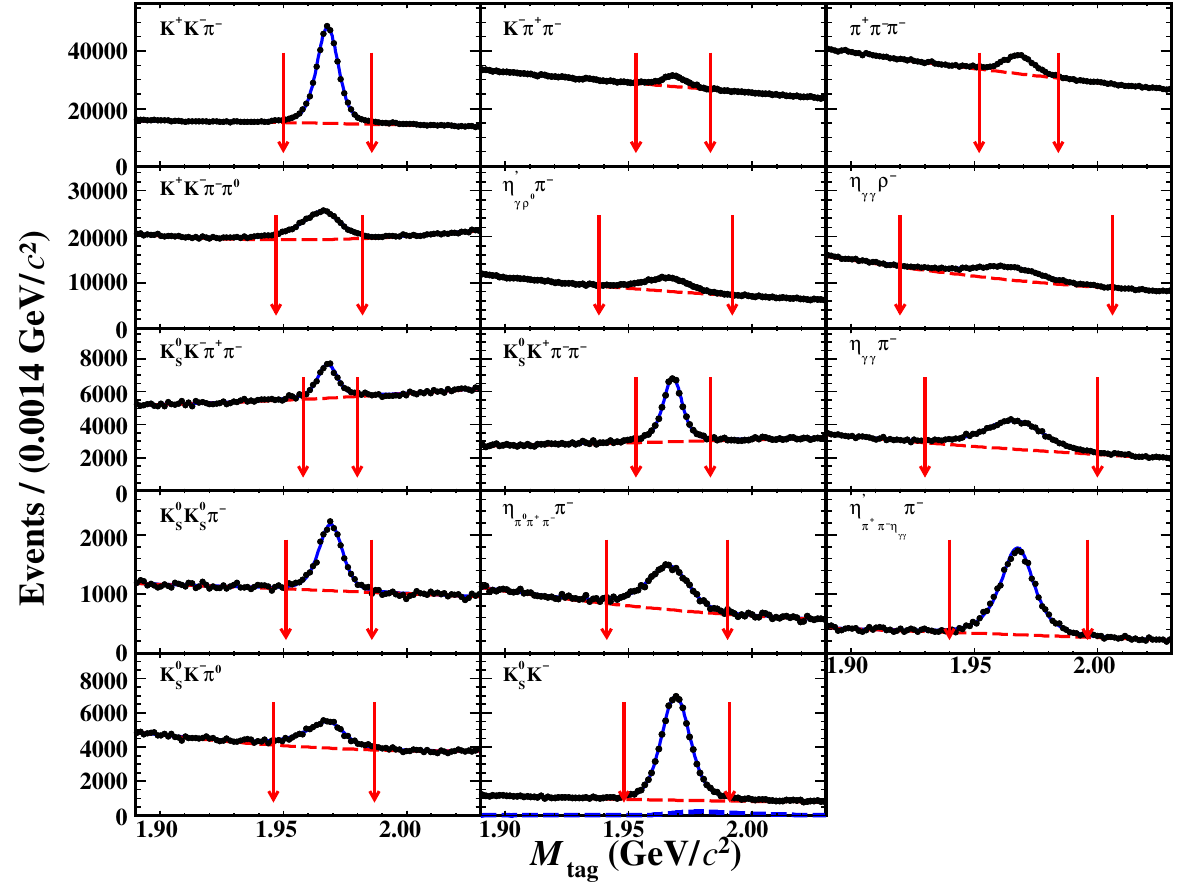}
\caption{\footnotesize
Fits to the $M_{\rm tag}$ distributions of the ST $D^-_s$  candidates.
Points with error bars are data for all energy points.
Blue solid lines are the best fits. Red dashed lines are the fitted background shapes.
For the  $K_S^0K^-$ tag mode, the blue doted line is from the  $D^-\to K_S^0\pi^-$ background.
For each tag mode, the $M_{\rm tag}$ signal region are denoted by the pair of red arrows.
}
\label{fig:stfit}
\end{figure*}

If there are multiple combinations in one  event, only the  candidate  with the $D_s^-$ recoil mass
\begin{equation}
M_{\rm rec} \equiv \sqrt{ \left (E_{\rm CM} - \sqrt{|\vec p_{D^-_s}|^2 +m^2_{D^-_s} } \right )^2
-|\vec p_{D^-_s}|^2}
\end{equation}
closest to the $D_s^{*+}$ nominal mass~\cite{ParticleDataGroup:2020ssz} is kept for further analysis per tag mode per charge. Here, $\vec{p}_{D_s^-}$ is the momentum of the $D^-_s$ candidate and $m_{D_s^{-}}$ is the nominal  $D_s^-$ meson~\cite{ParticleDataGroup:2020ssz}. Figure~\ref{fig:stfit} shows the invariant mass ($M_{\rm tag}$) spectra of the accepted ST $D^-_s$ candidates  for the individual tag modes in data combined from all energy points.  The ST yield for each tag mode is obtained through fitting the corresponding $M_{\rm tag}$ spectrum.  In the fit, the  signal is modeled  by the simulated  shape, for events where the solid angle between the generated and reconstructed four-momentum is no more  than $15^{\circ}$, convolved with a Gaussian function to take into account the resolution difference between data and simulation. 

For the  $D^-_s\to K_S^0K^-$ tag mode, there is a peaking background  from  $D^-\to K^0_S\pi^-$, which is modeled  by the simulated  shape convolved with the same Gaussian function used in the signal shape with its size left as a free parameter. A second-order polynomial is used to  describe  the non-peaking background, which has been validated with  the inclusive MC  sample. The fit results are shown  in figure~\ref{fig:stfit}. Events within the signal regions are kept for the  further analyses.  As an example, the ST yields ($N_{\alpha, 2}^{\text{ST}}$) for different tag modes in data at $E_{\rm CM}$ = 4.178~GeV and the corresponding ST and DT efficiencies ($\epsilon_{\alpha,2}^{\text{ST}}, \epsilon^{\text{DT}}_{\alpha,\text{sig},2}$) are summarized in  Table~\ref{tab:bf}.  The values of $N_{\alpha, j}^{\text{ST}}$ and $\epsilon_{\alpha,j}^{\text{ST}}$ at the other energy points are obtained similarly. The ST yields $N_{ j}^{\text{ST}}$  in data and  the averaged signal efficiencies $\bar{\epsilon}_{j}$ at each energy point  are summarized in  Table~\ref{tab:eff2}. Summing over all  tag modes and energy points gives the total ST yield to be   $N^{\rm tot}_{\rm ST}$= 816634±3679, where the uncertainty is statistical only.

\begin{table}[htbp]
\centering\linespread{1.15}
        \caption{
        The obtained values of $N_{\alpha, 2}^{\text{ST}}$, $\epsilon_{\alpha,2}^{\text{ST}}$, and $\epsilon^{\text{DT}}_{\alpha,\text{sig},2}$ in the $\alpha$ tag mode at $E_{\rm CM}=4.178$~GeV, where the efficiencies do not include the BFs for the intermediate resonance decays and the uncertainties are statistical only.}
\small
        \label{tab:bf}
        \begin{tabular}{lr@{}lr@{}lr@{}l}\hline
Tag mode  &\multicolumn{2}{c}{$N_{\alpha, 2}^{\text{ST}}$}&\multicolumn{2}{c}{$\epsilon_{\alpha,2}^{\text{ST}}$~(\%)}&\multicolumn{2}{c}{$\epsilon^{\text{DT}}_{\alpha,\text{sig},2}~(\%)$} \\ \hline
$K^{+}K^{-}\pi^{-}$&137317&$\pm$608&40.92&$\pm$0.02&6.84&$\pm$0.03\\
$K^-\pi^+\pi^-$&16514&$\pm$632&45.42&$\pm$0.07&8.52&$\pm$0.12\\
$\pi^{+}\pi^{-}\pi^{-}$&36497&$\pm$873&52.13&$\pm$0.05& 9.97&$\pm$0.09\\
$K^{+}K^{-}\pi^{-}\pi^{0}$&42119&$\pm$851&11.77&$\pm$0.01&2.03&$\pm$0.02\\
$\eta^{'}_{\gamma\rho^{0}}\pi^{-}$&24698&$\pm$656&32.53&$\pm$0.04&6.78&$\pm$0.08\\
$\eta_{\gamma\gamma}\rho^{-}$&39670&$\pm$1673&19.88&$\pm$0.02&4.99&$\pm$0.04\\
$K_{S}^{0}K^{-}\pi^{+}\pi^{-}$&7621&$\pm$270&18.51&$\pm$0.05& 3.07&$\pm$0.07\\
$K_{S}^{0}K^{+}\pi^{-}\pi^{-}$&14855&$\pm$235&21.12&$\pm$0.04&3.68&$\pm$0.06\\
$\eta_{\gamma\gamma}\pi^{-}$&19239&$\pm$468&48.79&$\pm$0.06&9.97&$\pm$0.12\\
$K_{S}^{0}K_{S}^{0}\pi^{-}$&5088&$\pm$149&22.82&$\pm$0.07& 4.20&$\pm$0.10\\
$\eta_{\pi^{+}\pi^{-}\pi^{0}}\pi^{-}$&5693&$\pm$201&23.49&$\pm$0.07&4.83&$\pm$ 0.11\\
$\eta^{'}_{\pi^{+}\pi^{-}\eta_{\gamma\gamma}}\pi^{-}$&9730&$\pm$140&25.26&$\pm$0.05& 4.89&$\pm$0.08\\
$K_{S}^{0}K^{-}\pi^{0}$&11182&$\pm$449&17.01&$\pm$0.04&3.64&$\pm$0.06\\
$K_{S}^{0}K^{-}$&30956&$\pm$261&47.63&$\pm$0.05&9.29&$\pm$0.09\\
\hline
        \end{tabular}
\end{table}

\begin{table}[htbp]\centering\linespread{1.15}
	\caption{
	The requirements on $M_{\rm BC}$, the total ST yields ($N_{j}^{\text{ST}}$), and the averaged signal efficiencies ($\bar{\epsilon}_{j}$ =$(\sum_{\alpha} (N_{\alpha, j}^{\text{ST}}/N_{j}^{\text{ST}})\cdot (\epsilon^{\text{DT}}_{\alpha,\text{sig},j}/\epsilon_{\alpha,j}^{\text{ST}})$) at various energy points, where the efficiencies do not include the BF of $\phi \to K^+K^-$ and the uncertainties are statistical only. The definition of $M_{\rm BC}$ is given in Eq.~(\ref{Mbc}).}
\small
	\label{tab:eff2}
	\begin{tabular}{l@{}cr@{}lr@{}lr}\hline
$E_{\rm CM}$ (GeV) & \multicolumn{1}{r}{$M_{\rm BC}$ ({GeV}/$c^2)$}&\multicolumn{2}{c}{$N_{j}^{\text{ST}}$}&\multicolumn{2}{c}{$\bar{\epsilon}_{j}~(\%)$} \\ \hline
4.128&[2.010,\,\,2.061]&31803&$\pm$695&18.07&$\pm$0.06\\
4.157&[2.010,\,\,2.070]&51026&$\pm$839&19.03&$\pm$0.06\\
4.178&[2.010,\,\,2.073]&401179&$\pm$2487&18.84&$\pm$0.05\\
4.189&[2.010,\,\,2.076]&67575&$\pm$927&19.97&$\pm$0.06\\
4.199&[2.010,\,\,2.079]&63076&$\pm$950&19.51&$\pm$0.06\\
4.209&[2.010,\,\,2.082]&63119&$\pm$1052&20.24&$\pm$0.06\\
4.219&[2.010,\,\,2.085]&53466&$\pm$943&20.46&$\pm$0.06\\
4.226&[2.010,\,\,2.088]&85390&$\pm$1551&21.64&$\pm$0.06\\
		\hline
	\end{tabular}
\end{table}

\section{Selection of $D_s^+\to K^+K^-\mu^+\nu_\mu$}
After the selection of the tagged $D_s^-$ candidate, a transition $\gamma$ or $\pi^0$ is searched for among the unused (by the ST) photon candidates passing the basic criteria mentioned before.  All possible $\gamma$ or $\pi^0$ candidates are looped over; if there are multiple candidates, 
the one giving the minimum $|\Delta E|$ is kept.  
Here, $|\Delta E|$ is defined as 
\begin{equation}
\Delta E \equiv E_{\rm CM} - E_{D^-_s} - \sqrt{  |-\vec{p}_{\gamma(\pi^0)}-\vec{p}_{D^-_s}|^2+m^2_{D^{+}_s}}
-E_{\gamma(\pi^0)}, 
\end{equation}
with $E_{\gamma, \pi^0, D^-_s } $ and $\vec{p}_{\gamma, \pi^0, D^-_s}$ being the respective energy and momentum.  
In the presence of the ST $D_s^-$ and transition $\gamma$($\pi^0$), the final state particles of the signal decay are selected from among the residual tracks. $K^+$ and $K^-$ candidates are selected in a similar manner as the ST decay products. The muon candidates are identified based on combined information of the $\mathrm{d}E/\mathrm{d}x$ measurement from the MDC, the TOF data and the energy deposit in the EMC. The  combined likelihoods ${\cal L}_{e}$, ${\cal L}_{\mu}$, and ${\cal L}_{K}$ for the electron, muon, and kaon hypotheses are calculated and the muon candidates are required to  satisfy ${\cal L}_{\mu} > {\cal L}_{K}$, ${\cal L}_{\mu} > {\cal L}_{e}$, and  ${\cal L}_{\mu} >0.001$.
Then, with  information of the ST side, two kaons, muon, and $\gamma$ ($\pi^0$),  ${ U_{\rm miss}} $ is defined as:
 \begin{equation}
{ U_{\rm miss}} \equiv \left(E_{\rm CM} - E_{D^-_s} - E_{\rm \gamma/\pi^0} - E_{K^{+} K^{-}} - E_{\mu^+}\right) - \left|\vec{p}_{D^-_s} + \vec{p}_{\gamma/\pi^0} + \vec{p}_{K^{+} K^{-}} +\vec{p}_{\mu^+}\right|  
\end{equation}
as a signal variable related to the missing neutrino. To improve the ${ U_{\rm miss}}$ resolution, the candidate tracks and the missing neutrino are subjected to a 3-constraint kinematic fit.  Energy and momentum conservation along with three mass constraints is applied while the neutrino four-vector is determined.  The invariant masses of the two $D_s$ mesons are constrained to the nominal $D_s$ mass.  Finally, the invariant mass of the $D_s^-\gamma(\pi^0)$ or $D_s^+\gamma(\pi^0)$ combinations are constrained to the nominal $D_s^*$ mass, and the $D_s\gamma(\pi^0)$ combination with the smaller $\chi^{2}_{3C}$ is kept.
 
 It is required that both the number of unused reconstructed charged tracks ($N_{\rm extra}^{\rm char}$) and unused $\pi^0$ ($N_{\rm extra}^{\pi^0}$) candidates should be zero in all DT candidate events. To suppress the peaking background of $D_s^+\to K^{+} K^{-}\pi^+$, which is  caused by misidentifying a $\pi^+$ as a $\mu^+$, $M_{K^{+} K^{-}\nu_{\mu}}$ is required to be greater than 1.30 GeV/$c^2$. To reject the remaining background of $D^+_s\to K^{+} K^{-}\pi^+$, it is also required that $M_{K^{+} K^{-}\mu^+}$ is less than 1.75  GeV/$c^2$. To suppress the peaking background of $D_s^+\to K^{+} K^{-}\pi^+\pi^0$, which is mainly caused by misidentifying a the $\pi^+$ as a $\mu^+$ and missing the $\pi^0$, the maximal energy of the photons ($E_{\gamma  ~\rm extra }^{\rm max}$) is required to be less than 0.2 GeV. All requirements are obtained by optimizing the figure of merit  defined by $S/\sqrt{S+B}$, where $S$ and $B$ denote the signal and background yields based on normalized inclusive MC samples.  The optimizations of all requirements have been iterated for several times to obtain stable cuts.
 
  \section{Branching fraction measurement  }
\begin{figure}[htbp]
\centering
{\includegraphics[width=12cm]{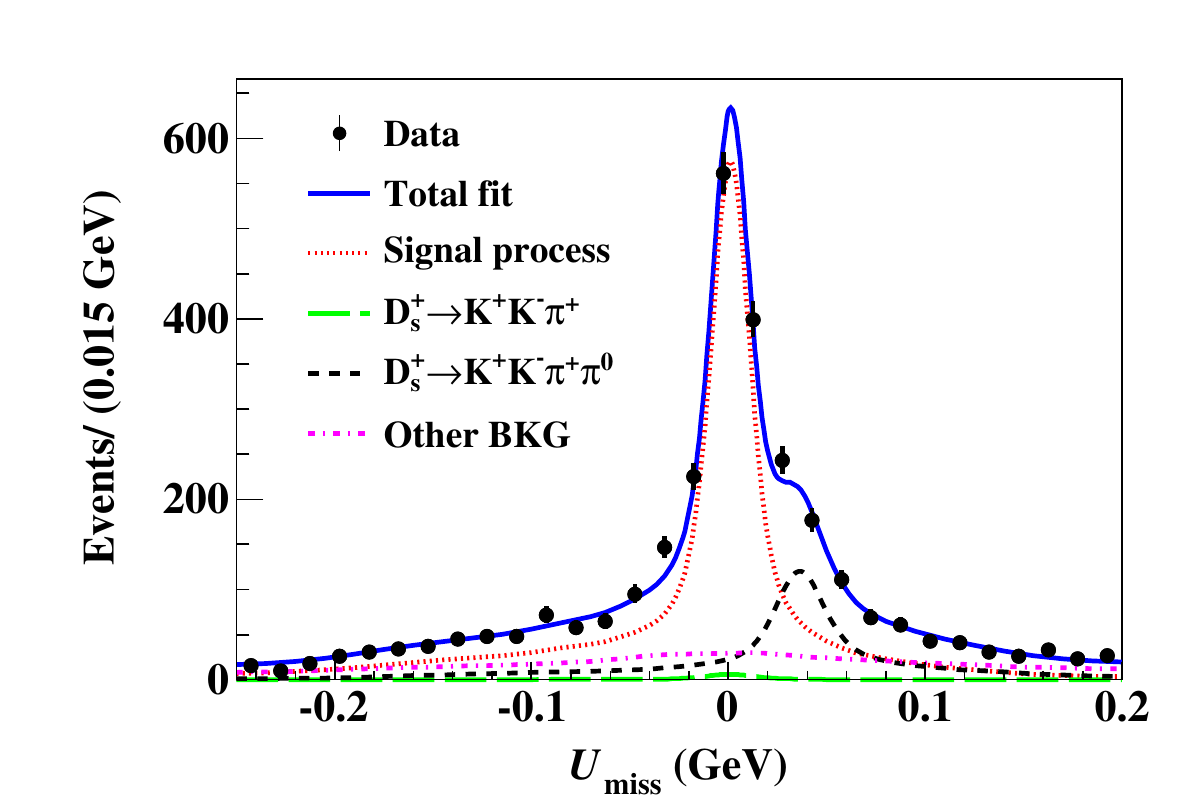}}

\caption{Fit to the ${ U_{\rm miss}}$ distribution of the candidate events for $\kkmv$. Points with error bars represent data. The blue solid curve denotes the total fit. The red dotted curve shows the signal process. Green long-dashed, black  short-dashed, and  dash-dotted violet curve  are the background contributions from $D_s^+\to K^+K^-\pi^+$, $D_s^+\to K^+K^-\pi^+\pi^0$,  and  the other background contributions, respectively.
}
\label{fig:fit_ep}
\end{figure}

 After imposing all of the aforementioned selection criteria, the resulting  ${ U_{\rm miss}}$ distribution of the accepted $D_s^+\to K^+K^-\mu^+\nu_\mu$ candidate events in the data sample is shown in figure~\ref{fig:fit_ep}. To extract the signal yield, an unbinned extended maximum likelihood fit is performed on this distribution. The signal is modeled by the MC-simulated shape convolved with a Gaussian function, which represents the resolution difference between data and MC samples. The peaking background of  $D_s^+\to K^+K^-\pi^+$  is  fixed according to the MC simulations and the peaking background of $D_s^+\to K^+K^-\pi^+\pi^0$  is allowed to float.  Other backgrounds are dominated by  processes of open charm production and continuum $q\bar q$, which are modeled by the inclusive MC simulation with a luminosity about 40 times that of the data sample.  
 
 The PWA (described later) shows  that the only significant resonance contribution to the $K^+K^-$ system is $\phimv$. 
  Using  Eq.~(\ref{eq1}), where the ${\mathcal B}_{\phi \to K^+K^- }= (49.1\pm 0.5) \times10^{-2}$~\cite{ParticleDataGroup:2020ssz} and the signal yield is $1725\pm68$, the ST yield and the ST/DT efficiencies are mentioned  in Sec.~\ref{ST-selection}, the corresponding BF is determined to be  ${\mathcal B}(\phimv)= (2.25\pm 0.09 \pm 0.07) \times10^{-2}$, where the first uncertainty is statistical, while the second is systematic. The systematic uncertainties will be described in more detail in Sec.~\ref{sec: Systematic Uncertainties}.

\section{Partial wave analysis}
\label{sec:Partial Wave Analysis}
 To obtain data samples with high purity for the PWA, two further constraints of $|{ U_{\rm miss}}|<0.02$ GeV and $\chi^{2}_{3C}< 100$  are imposed on the accepted candidates. 
  After these conditions, 939 signal events remain, with an estimated average background level of (9.8 $\pm$ 0.7)\%  at all energy points. 
\subsection{Kinematics and decay rate formalism}
The differential decay rate for $D^+_s \to K^+ K^- \mu^+ \nu_\mu$ depends on five variables \cite{Cabibbo:1965zzb}:
$m$, the invariant mass of the $K^+ K^-$ system;
$q$, the invariant mass of the $\mu^+ \nu_\mu$ system;
$\theta_\mu$ ($\theta_{K}$), the angle between the momentum of the $\mu^+$ ($K^-$) in the $\mu^+ \nu_\mu$($K^+ K^-$) rest frame and
the momentum of the $\mu^+ \nu_\mu$ ($K^+ K^-$) system in the $D^+_s$ rest frame;
and $\chi$, the angle between the normals of the decay planes defined in the $D^+_s$ rest frame by the $K^+ K^-$ pair and the $\mu^+ \nu_\mu$ pair.

The differential decay rate as a function of these variables is given in Ref.~\cite{Cabibbo:1965zzb}.
The formula is updated in Ref.~\cite{Lee:1992ih}, based on chiral perturbation theory and heavy-quark symmetry; muon  mass effects are treated in Ref.~\cite{Zhang:2023nnn}.  
The differential decay width of $D^+_s \to K^+ K^- \mu^+ \nu_\mu$ is expressed as

\begin{equation}
  \begin{aligned}
  d^{5}\Gamma &= \frac{G^{2}_{F}|V_{cs}|^{2}}{(4\pi)^{6}m^{3}_{D_s}}X\beta_{m } \beta_{l } \, {\cal I}(m^{2}, q^{2}, \theta_{K^-}, \theta_{\mu}, \chi) \,  dm^{2} dq^{2} d{\rm cos} \theta_{K^-} {\it d}{\rm cos}\theta_{\mu}d\chi, 
  \label{eq:decay_rate}
  \end{aligned}
\end{equation}
where $X=p_{K^+K^-}m_{D_s}$, $p_{K^+K^-}$ is the modulus of the momentum of the $K^+K^-$ in the $D^+_s$ rest frame,  $\beta_{m}=2p^{*}/m$ and $\beta_{l}=2p^{\prime}/q$, in which $p^{*}$ is the modulus of the momentum of the $K^-$ in the $K^+K^-$ rest frame, while $p^{\prime}$ is defined as  the modulus of the momentum of $\mu^+$ in the $\mu^+\nu_{\mu}$ rest frame.
The Fermi coupling constant is denoted by $G_{F}$.
The  decay density $\cal I$ is given by
\begin{equation}
  \begin{aligned}
  {\cal I} = & {\cal I}_{1}+{\cal I}_{2}{\rm cos2}\theta_{\mu}+{\cal I}_{3}{\rm sin}^{2}\theta_{\mu}{\rm cos}2\chi+{\cal I}_{4}
  {\rm sin2}\theta_{\mu}{\rm cos}\chi+{\cal I}_{5}{\rm  sin}\theta_{\mu}{\rm cos}\chi  \\
  &+{\cal I}_{6}{\rm  cos}\theta_{\mu}+{\cal I}_{7}{\rm sin}\theta_{\mu}{\rm sin}\chi+{\cal I}_{8}{\rm sin2}\theta_{\mu}{\rm sin}\chi
  +{\cal I}_{9}{\rm sin}^{2}\theta_{\mu}{\rm sin2}\chi,
  \label{eq:decay_intensity}
  \end{aligned}
\end{equation}  
where ${\cal I}_{1, \ldots,9}$ depend on $m^{2}$, $q^{2}$, and $\theta_{K^-}$. These quantities can be expressed in terms of the four FFs ${\cal F}_{1,2,3,4}$. 
Then one can expand ${\cal F}_{i=1,2,3,4}$ into partial wave amplitudes including $S$-wave (${\cal F}_{10}$), $P$-wave (${\cal F}_{i1}$), and $D$-wave (${\cal F}_{i2}$),
%The ${\cal F}_{i=1,2,3,4}$  can be expanded in partial wave amplitudes:  $S$-wave (${\cal F}_{10}$), $P$-wave (${\cal F}_{i1}$), and $D$-wave (${\cal F}_{i2}$),
to show their explicit dependence on $\theta_{K^-}$. The detailed formulas can  be found  in Ref.~\cite{Zhang:2023nnn}.
Based on the existing data, we do not find $D$-wave components, so the amplitude ${\cal F}_{i2}$   is ignored.
 Consequently, the FFs can be written as
\begin{equation}
  \begin{aligned}
  {\cal F}_{1}&={\cal F}_{10}+{\cal F}_{11}cos\theta_{K^-},~  
  {\cal F}_{2}&=\frac{1}{\sqrt{2}}{\cal F}_{21}, ~
  {\cal F}_{3}&=\frac{1}{\sqrt{2}}{\cal F}_{31}, ~
    {\cal F}_{4}&={\cal F}_{41}cos\theta_{K^-},~
  \label{eq:form_factor}
  \end{aligned}
\end{equation}
where  ${\cal F}_{i1}$ can be parameterized with the helicity basis FFs $H_{0,\pm}(q^{2})$.
 The helicity FFs can in turn be related to two axial-vector FFs $A_{1,2}(q^{2})$ and one vector FF $V(q^{2})$. The $A_{1,2}(q^{2})$ and $V(q^{2})$ all take the simple pole form $A_{i}(q^{2})={A_{1,2}(0)}/({1-{q^{2}}/{m_{A}^{2}}})$ and $V(q^{2})={V(0)}/({1-{q^{2}}/{m_{V}^{2}}})$,  and 
the pole mass $m_{V}$ and $m_{A}$ are fixed to $m_{D^*_s} \simeq $ 2.1 GeV/$c^{2}$ and $m_{D_{s1}} \simeq$ 2.5~GeV/$c^{2}$, respectively.
The FF $A_{1}(q^{2})$ is common to all three helicity amplitudes. Therefore, it is natural to define the two coupling constants, $r_{V}=V(0)/A_{1}(0)$ and $r_{2}=A_{2}(0)/A_{1}(0)$ as FF ratios at the momentum square $q^{2}=0$. They are determined from the PWA fit.

The amplitude of the $P$-wave resonance $\mathcal{A}(m)$ is expressed as a relativistic Breit-Wigner
\begin{equation}
\mathcal{A}(m)= \frac{m_{0}\Gamma_{0}(p^{*}/{p^{*}_{0})}}{m_{0}^{2}-m^{2}-im_{0}\Gamma(m)}\frac{B(p^{*})}{B(p^{*}_{0})}, 
\label{eq:Am}
\end{equation}
where $B(p) = 1/\sqrt{1+r_{BW}^{2}p^{2}}$ with $r_{BW}$ = 3.0 $({\rm GeV}/c)^{-1}$ and $\Gamma(m)=\Gamma_{0} \left(\frac{p^{*}}{p^{*}_{0}}\right)^{3}\left(\frac{m_{0}}{m}\right)\left[\frac{B(p^{*})}{B(p^{*}_{0})}\right]^{2}$, where   $p^{*}_{0}$ is the modulus of the momentum of the $K^-$  at the pole mass of the resonance $m_0$.

The $S$-wave contribution, characterized by the FF ${\cal F}_{10}$, is parametrized, assuming only $f_0(980)$ production, as 

\begin{equation}
  {\cal F}_{10}=p_{K^+K^-}m_{D_s}\frac{1}{1-\frac{q^{2}}{m_{A}^{2}}}\mathcal{A}_{S}(m),
\label{eq:form_factor_S}
\end{equation}
where $p_{K^+K^-}$ is the modulus of the momentum of the $K^+K^-$ system in the $D^+_s$ rest frame. Here the term $\mathcal{A}_{S}(m)$ corresponds to the mass-dependent $S$-wave amplitude.  The  Flatt\'e formula is used for the $f_0(980)$ contribution,

\begin{equation}
\mathcal{A}_{S}(m)=\frac{a_{S}e^{i\delta_S}}{m^2_0-m^2-i(g_1\rho_{\pi\pi}+g_2\rho_{KK})},  
\label{eq:flatte}
\end{equation} 
where $a_{S}$ and $\delta_S$ are the magnitude and phase of the S-wave amplitude; they are relative to the $D^+_s\to \phi\mu^+\nu_\mu$ amplitude. 
The parameters $g_1$ and $g_2$ are taken from Ref.~\cite{BES:2004twe};
 $\rho_{\pi\pi}$ and  $\rho_{KK}$ are the phase-space (PHSP) factors for the decay channels $\pi\pi$ and $KK$, respectively.
\subsection{Fit method}
The  PWA fit is performed using an unbinned maximum likelihood method.
For one candidate event, the probability density function ($PDF$) can be expressed as:
\begin{equation}
  PDF(\xi,\eta)=(1-f_{b}) \mathcal{S} + f_{b}\mathcal{B}=(1-f_{b}) \frac{\omega(\xi,\eta) \, \epsilon(\xi)}{\int d\xi \, \omega(\xi,\eta) \, \epsilon(\xi)} +f_{b}\frac{B(\xi)}{\int d\xi \, B(\xi)},
  \label{eq:pdf1}
\end{equation}
where $\xi$ denotes the five kinematic variables characterizing of one event  and $\eta$ denotes the fit parameters  such as $r_V$ and $r_2$;
$\omega(\xi,\eta)$ is the decay intensity, and $B(\xi)$ is a function that describes the background; $\epsilon(\xi)$ is the reconstruction efficiency for the final state $\xi$ and $f_{b}$ is the fraction of background events.  
The above PDF can be rewritten as: 
\begin{equation}
  PDF(\xi,\eta)=(1-f_{b}) \mathcal{S} + f_{b}\mathcal{B}=\epsilon(\xi)\left[(1-f_{b}) \frac{\omega(\xi,\eta)}{\int d\xi \, \omega(\xi,\eta) \, \epsilon(\xi)} +f_{b}\frac{B_{\epsilon}(\xi)}{\int d\xi \, B_{\epsilon}(\xi) \, \epsilon(\xi)}\right],
  \label{eq:pdf2}
\end{equation}
where $B_{\epsilon}(\xi)$ is defined to be the background distribution corrected by the acceptance function $\epsilon(\xi)$~\cite{CLEO:2012beo}. 
By factorizing $\epsilon(\xi)$ out as a common factor, it becomes a part of the normalization. Then the likelihood is the product of probabilities of all the events:
\begin{equation}
  {\cal L}=\prod_{i=1}^{N}PDF(\xi_{i}, \eta)
   =\prod_{i=1}^{N} \epsilon(\xi_{i})\left[(1-f_{b}) \frac{\omega(\xi_{i},\eta)}{\int d\xi_{i} \, \omega(\xi_{i},\eta) \, \epsilon(\xi_{i})} +f_{b}\frac{B_{\epsilon}(\xi_{i})}{\int d\xi_{i} \, B_{\epsilon}(\xi_{i}) \, \epsilon(\xi_{i})}\right].
\end{equation}
\noindent 
In the fit, we optimize the parameters $\eta$ by 
performing a minimization of a negative log-likelihood ($NLL$): 
\begin{equation} 
  -\ln\!{\cal L} =
  -\sum_{i=1}^{N}\ln(\epsilon(\xi_{i})) 
  -\sum_{i=1}^{N}\ln\left[(1-f_{b}) \frac{\omega(\xi_{i},\eta)}{\int d\xi_{i} \, \omega(\xi_{i},\eta) \, \epsilon(\xi_{i})} +f_{b}\frac{B_{\epsilon}(\xi_{i})}{\int d\xi_{i} \, B_{\epsilon}(\xi_{i}) \, \epsilon(\xi_{i})}\right].
  \label{eq:lnL}
\end{equation}
\noindent 
The first term in Eq.~\ref{eq:lnL} depends only on the events and efficiency, 
and remains constant during the fit. So actually we only compute the second term while performing the fit.
Let $\sigma_S$ be $\int d\xi_{i}\omega(\xi_{i},\eta)\epsilon(\xi_i)$ and $\sigma_B$ be $\int d\xi_{i} B_{\epsilon}(\xi_{i})\epsilon(\xi_{i})$.
We minimize $NLL$:
\begin{equation} 
  NLL=-\sum_{i=1}^{N}\ln\left[(1-f_{b})\frac{\omega(\xi_{i},\eta)}{\sigma_S}+f_{b}\frac{B_{\epsilon}(\xi)}{\sigma_B}\right]. 
  \label{eq:nll}
\end{equation}
\noindent 
The acceptance efficiency has been considered in the calculation 
of the normalization integral factors $\sigma_S$ and $\sigma_B$, 
which we calculate with MC integration using the signal MC.
The normalization integral terms can be given as:
\begin{eqnarray}
\sigma_S=\int d\xi_{i} \, \omega(\xi_{i},\eta) \, \epsilon(\xi_i)
      \propto\frac{1}{N_{\rm selected}}\sum_{k=1}^{N_{\rm selected}} \frac{\omega(\xi_{k}, \eta)}{\omega(\xi_{k}, \eta_{0})},
\label{eq:sigmc-integral}
\end{eqnarray}

\begin{eqnarray}
\sigma_B=\int d\xi_{i} \, B_{\epsilon}(\xi_{i}) \, \epsilon(\xi_{i})
      \propto\frac{1}{N_{\rm selected}}\sum_{k=1}^{N_{\rm selected}} \frac{ B_{\epsilon}(\xi_{k})}{\omega(\xi_{k}, \eta_{0})}.
\label{eq:sigmc-integral}
\end{eqnarray}
Here the terms $\eta$ and $\eta_{0}$ represent the values of the parameters used in the fit 
and those used to produce the simulated events, respectively, while
$N_{\rm selected}$ denotes the number of the signal MC events after reconstruction and selection. 

 A correction $\gamma_\epsilon$ is introduced to account for the potential bias caused by the tracking and PID efficiency differences between data and MC simulations.
By weighting each signal MC event with $\gamma_{\epsilon}$,  the MC integration is given by
\begin{eqnarray}
\sigma_S=\int d\xi_{i} \, \omega(\xi_{i},\eta) \, \epsilon(\xi_i)
      \propto\frac{1}{N_{\rm selected}}\sum_{k=1}^{N_{\rm selected}} \frac{\omega(\xi_{k}, \eta) \, \gamma_\epsilon(\xi_{k})}{\omega(\xi_{k}, \eta_{0})}.
\label{neweq:sigmc-integral}
\end{eqnarray}

The background shape is modeled with the inclusive MC and its fraction $f_b$ is fixed according to the result of the $U_{\rm miss}$ fit.
We model the background with non-parametric functions belonging to the class RooNDKeysPDF that use an adaptive kernel-estimation algorithm~\cite{Cranmer:2000du}. The value of $\epsilon(\xi_i)$ is obtained by calculating the ratio between the numbers of selected and truth events using PHSP MC samples, which are divided into 3 × 5 × 4 × 5 × 3 bins in the five-variable space $(m^{2}, q^{2}, \theta_{K^-}, \theta_{\mu}, \chi)$. For some edge bins, we merge neighboring bins until twenty events are accumulated. 

The data samples from the entire energy interval [4.128, 4.226] GeV are divided into two groups: one group is 4.178 GeV, while the other combines the intervals [4.128, 4.157] GeV and [4.189, 4.226] GeV.  
The reason for combining the latter intervals is their low statistics.
A simultaneous fit is performed to the two groups with 
the combined likelihood function: 
\begin{equation}
 {\cal L}^{ab}={\cal L}^{a}(\xi|\eta){\cal L}^{b}(\xi|\eta)=\prod_{i=1}^{n}PDF^{a}(\xi_i|\eta)\prod_{j=1}^{m}PDF^{b}(\xi_j|\eta),
\end{equation}
where $\xi$ and $\eta$ are defined as before and $a, b$ denote the likelihood values for the two data groups mentioned above.

\subsection{PWA results}
\begin{figure}[htp]
  \centering
\includegraphics[width=0.32\textwidth]{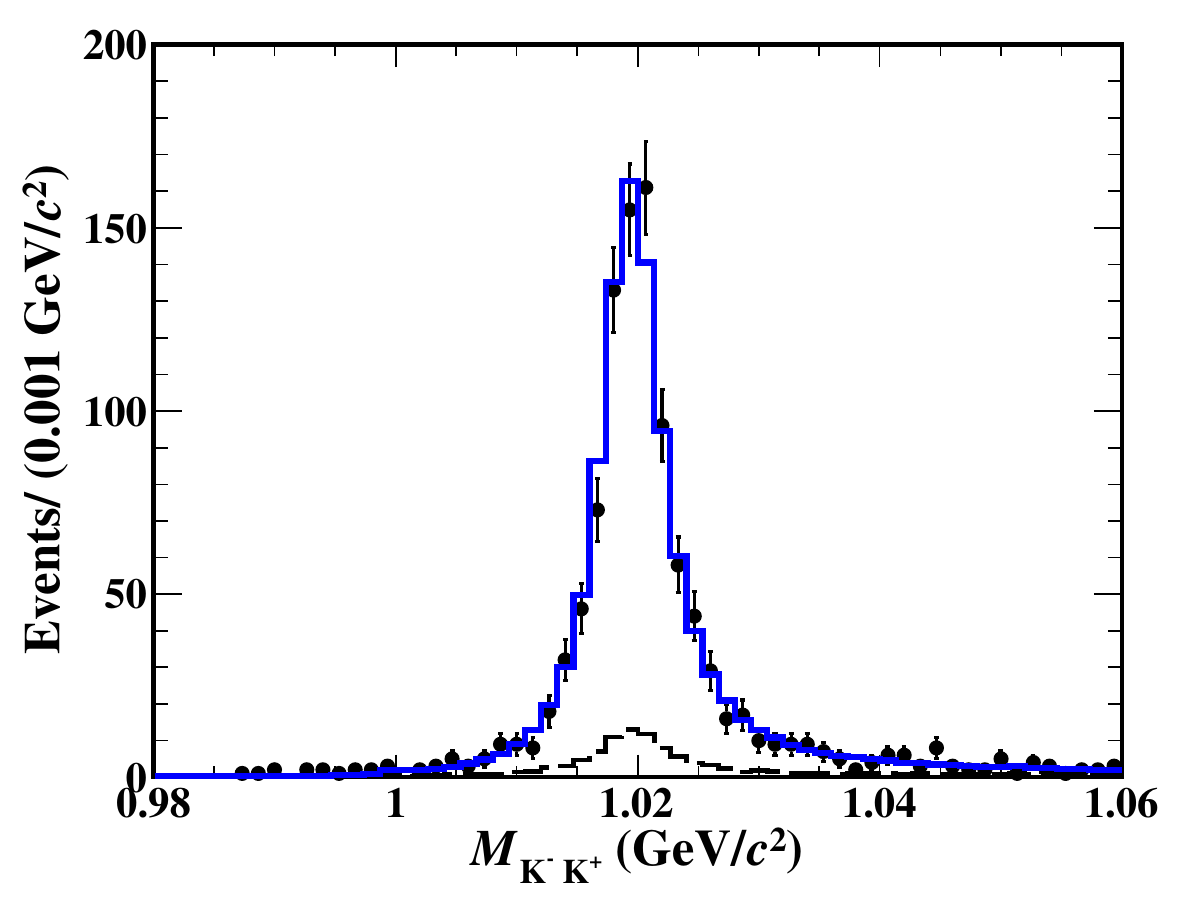}
\includegraphics[width=0.32\textwidth]{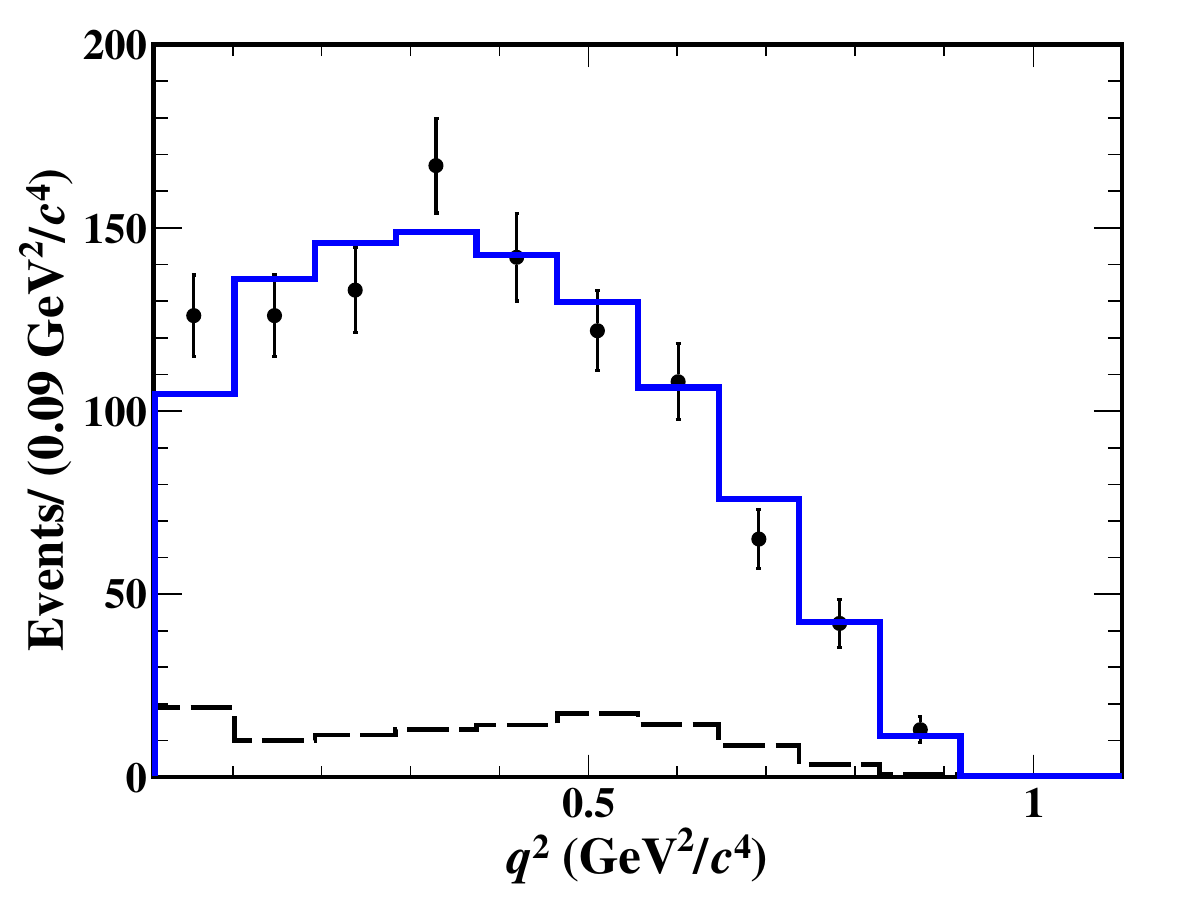} \\
\includegraphics[width=0.32\textwidth]{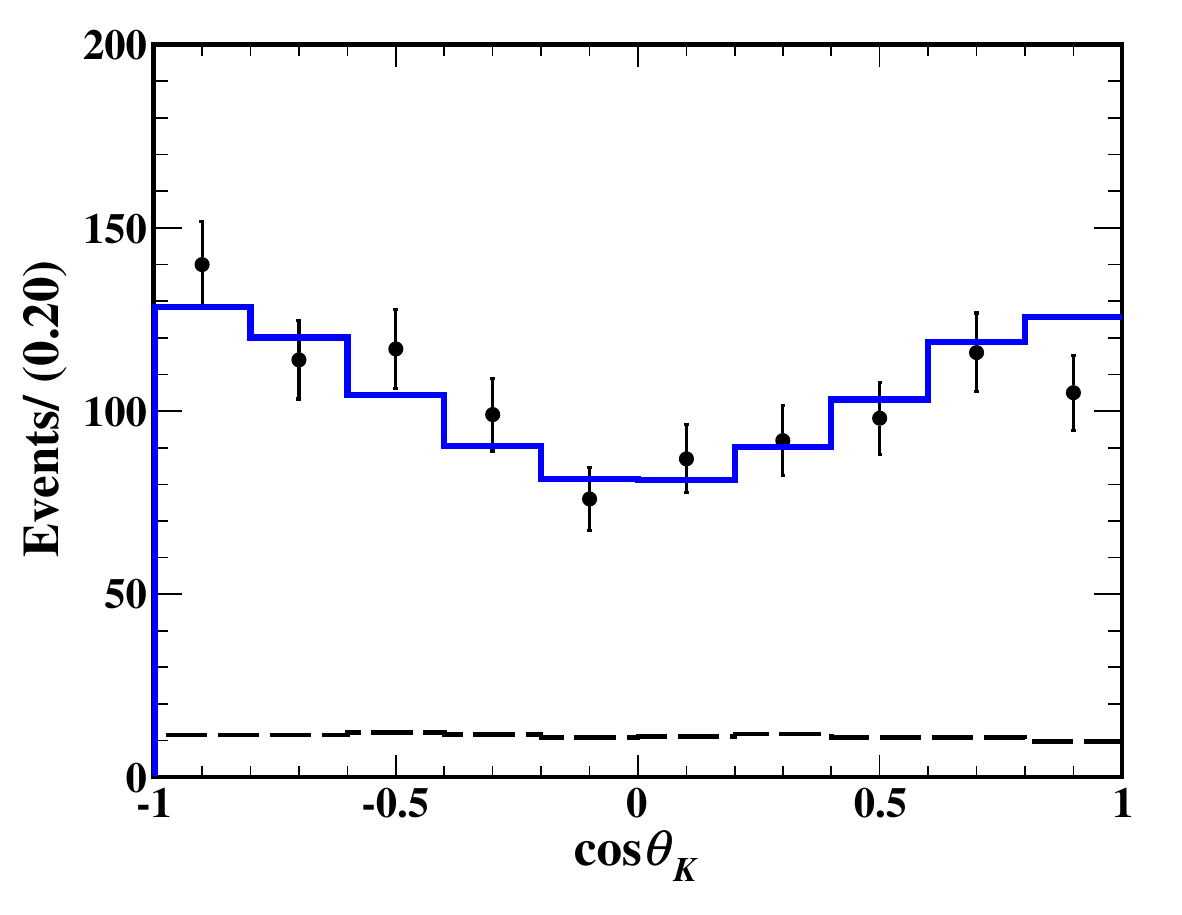}
\includegraphics[width=0.32\textwidth]{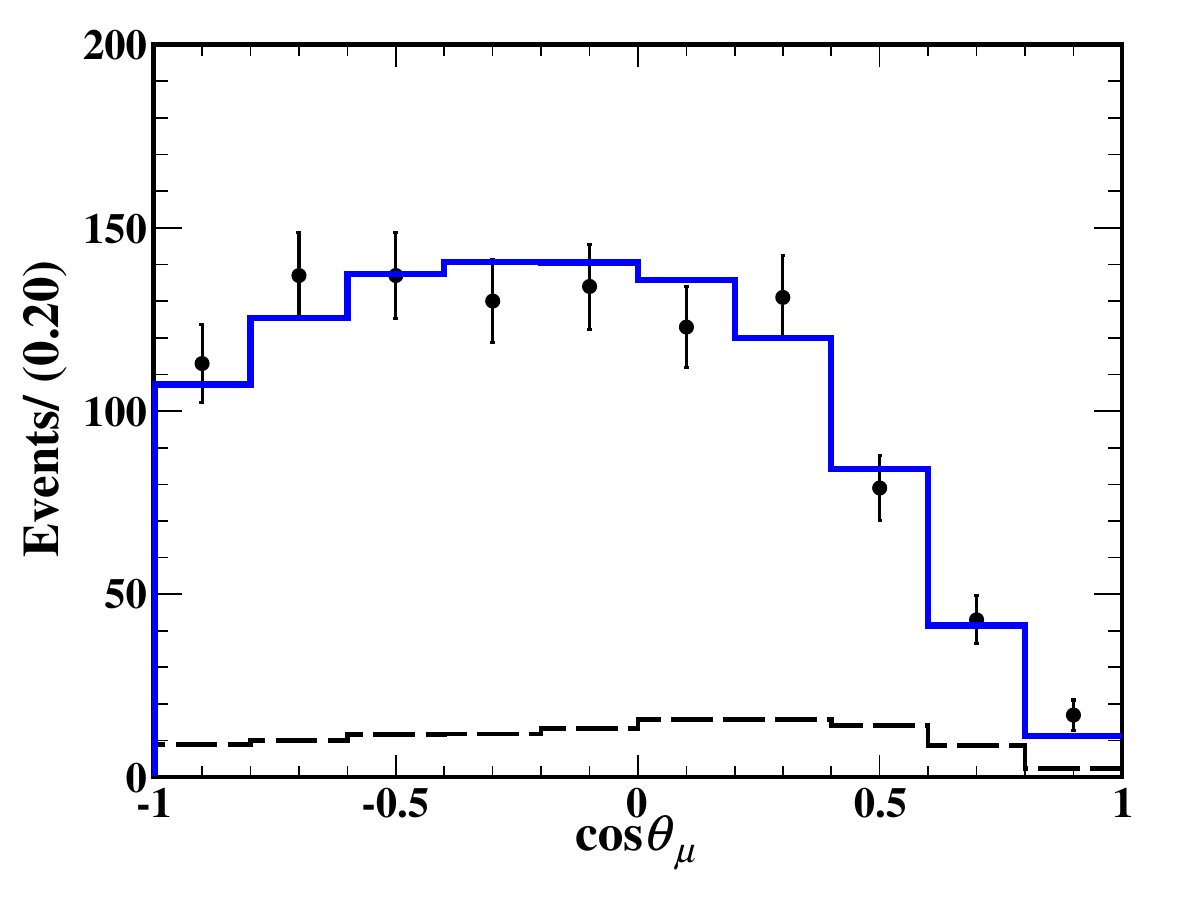}
\includegraphics[width=0.32\textwidth]{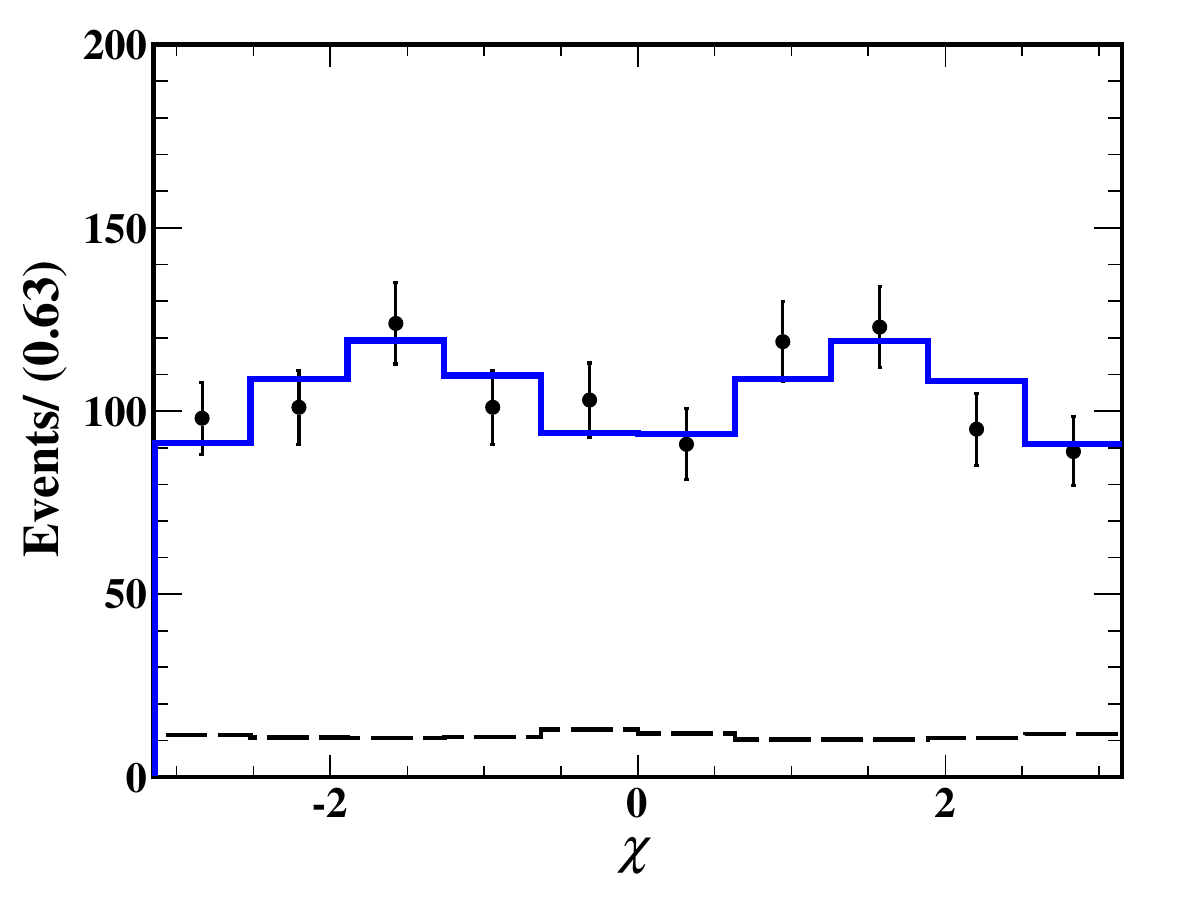}\\
\caption{Projections of the data and simultaneous PWA fit onto the five kinematic variables for  $D^+_s\to K^+K^-\mu^+\nu_\mu$.
The dots with error bars  are data, the blue lines are the best fit, and the dashed lines show the sum of the simulated background contributions.
}
\label{fit:pwa-center}
\end{figure}

A simultaneous PWA fit is performed on the two data groups. The structure of the $K^+K^-$ system is dominated by the vector meson $\phi$; nevertheless, the $S$-wave contribution from the $f_0(980)$  has been considered but no significant contribution has been observed. Therefore, only the $\phi$ in the $K^+K^-$ system is considered in the nominal solution. In the fit, the mass and width of $\phi$ are fixed to the PDG values~\cite{ParticleDataGroup:2020ssz}. The FF ratios $r_{V}=\frac{V(0)}{A_{1}(0)}=1.58\pm0.17\pm0.02$ and $r_{2}=\frac{A_{2}(0)}{A_{1}(0)}=0.71\pm0.14\pm0.02$ are obtained, with a correlation coefficient $\rho_{r_V,r_2}=-0.29$, where the first uncertainties are statistical and the second ones are systematic, see Sec.~\ref{sec: Systematic Uncertainties} for their derivation. The projections of the five kinematic variables  for the data are shown in figure~\ref{fit:pwa-center}. 

\begin{figure}[htp]\centering
\includegraphics[width=9cm]{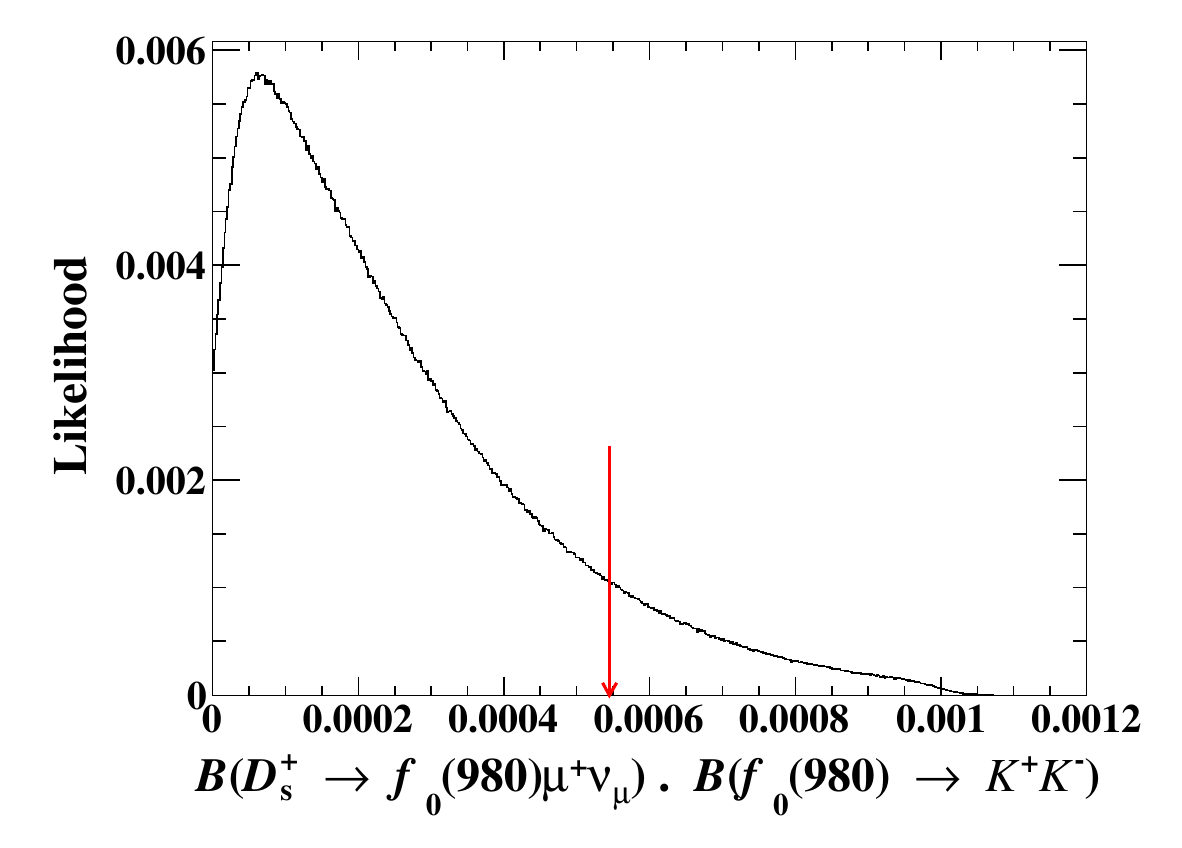}
\caption{Likelihood versus the product of ${\mathcal B}(D_s^+\to f_0(980)\mu^{+}{\nu}_{\mu}) \cdot {\mathcal B}(f_0(980)\to K^+K^-)$. The red arrow shows the upper limit at 90\% confidence level.
}
\label{upperlimit}
\end{figure}

A possible $f_0(980)$ component, an $S$-wave contribution to the ${\cal F}_{10}$ term, is studied by adding it to the nominal solution, where the $f_0(980)$ is parameterized with the Flatt\'e formula and the parameters are fixed based on the BES measurement~\cite{BES:2004twe}. The statistical significance of this component is only $2.2\sigma$ as determined by the change of $-2\ln\!{\mathcal L}$ in the PWA fits with and without this component, taking into account the change of the number of degrees of freedom.  
The systematic uncertainty on the $f_0(980)$ is estimated as $0.28\%$, similar to the central value. By scanning the magnitude of the $f_0(980)$ component, the likelihood variation of the expected BF is obtained and shown in figure~\ref{upperlimit}. To take the systematic uncertainty into account, the likelihood is convolved with a Gaussian function with a width equal to the systematic uncertainty.  The upper limit obtained is $\mathcal{B}(D_s^+\to f_0(980)\mu^{+}{\nu}_{\mu}) \cdot {\mathcal B}( f_0(980)\to K^+K^-) < 5.45 \times 10^{-4} $  at 90\% confidence level.

\section{ Systematic Uncertainties}
\label{sec: Systematic Uncertainties}
\subsection{Branching fraction measurement }
The sources of systematic uncertainty in the BF measurement are discussed below. 
\begin{itemize}
\item $N^{\rm tot}_{\rm ST}$.  The uncertainty due to $N^{\rm tot}_{\rm ST}$ is mainly from  the fits to the $M_{\rm tag}$ spectra. It is estimated by varying the signal and background shapes in the fits to data and inclusive MC sample. The alternative  signal shape is obtained  by varying  the matching requirement between generated and reconstructed angles from $15^{\circ}$ to $10^{\circ}$ and $20^{\circ}$. The alternative  background shape is obtained by replacing the nominal shape to be  a third order polynomial function. The difference of the ST total efficiency-corrected yields in data is taken as  systematic uncertainty. The uncertainty arising from the background fluctuation of the total ST yield is considered as a  systematic uncertainty.  Adding these three terms in quadrature gives the systematic uncertainty in $N^{\rm tot}_{\rm ST}$ to be 0.5\%.

\item $K^\pm$ tracking/PID  efficiency. The uncertainties in the tracking and  PID efficiencies of $K^\pm$  are studied with a control sample of $e^+e^-\to K^+K^-\pi^+\pi^-$.  The momentum-weighted data-MC differences are 1.008 $\pm$ 0.009 (1.009 $\pm$ 0.009) and  1.001 $\pm$ 0.004 (0.998 $\pm$ 0.004)  arising from  $K^+$ ($K^-$) tracking and PID efficiencies, respectively. The signal efficiencies applied to data are corrected by these factors. The uncertainties on these corrections are taken as the systematic uncertainties due to $K^\pm$ tracking and $K^\pm$ PID, as listed in Table~\ref{tab:bf-syst-sum}.

  \item $\mu^+$ tracking/PID  efficiency. The $\mu^+$ tracking and PID efficiencies are studied with a control sample of  $e^+e^-\to \gamma\mu^+\mu^-$.  The data-MC differences are 0.987 $\pm$ 0.003 for $\mu^+$ tracking and 1.040  $\pm$  0.004 for $\mu^+$ PID efficiencies. We correct the signal efficiencies to data  by these factors. The uncertainties on these corrections are taken as the systematic uncertainties due to $\mu^+$ tracking and $\mu^+$ PID, as listed in Table~\ref{tab:bf-syst-sum}.
  
\item  Transition $\gamma(\pi^0)$ reconstruction.   The efficiencies of the  $\gamma(\pi^0)$ reconstruction have been investigated with the control samples  of $J/\psi \to \pi^+\pi^-\pi^0$ ($e^+e^- \to K^+K^-\pi^+\pi^-\pi^0$). The systematic uncertainty of the $\gamma(\pi^0)$ selection is assigned to be 1.0\% in this analysis. 

\item Tag bias.  The ST efficiencies determined from the inclusive MC sample and those from the signal MC sample may be different, which may cause an uncertainty  associated with the ST selection, called tag bias. With the tracking and PID efficiencies for kaons and pions with different track multiplicities, the average difference, 0.1\%, is assigned as the systematic uncertainty due to tag bias.

\item  $E^{\rm max}_{\rm extra \gamma}$,  $N_{\rm extra}^{\rm char}$ and $N_{\rm extra}^{\pi^0}$.  The systematic uncertainty in the $E_{\rm extra \gamma}^{\rm max}$, $N_{\rm extra}^{\rm char}$, and $N_{\rm extra}^{\rm \pi^0}$  is estimated to be 0.4\% with the DT hadronic sample of  $D_s^+\to K^{+}K^{-}\pi^+$, 
a mode with tracks similar to the signal decay.  
The differences of the efficiencies between the data and MC simulation is $1.035 \pm 0.004$. 
After correcting the MC efficiency by this factor, we take 0.4\% as the systematic uncertainty.  

\item $M_{\phi \nu_{\mu}}$ and $M_{\phi \mu^+}$.   The uncertainties of the $M_{\rm \phi\nu_{\mu}}$ and $M_{\phi\mu^+}$ requirements are estimated with a control sample of $D_s^+\to \phi e^+\nu_e$, and the difference of the efficiencies between the data and MC simulation of 1.0\% is  taken as the uncertainty.  

\item  ${ U_{\rm miss}}$ fit. 
The uncertainty related to the ${U_{\rm miss}}$ fit is estimated with alternative signal and background shapes.  The systematic uncertainty from the  signal shape  is estimated by varying the $\rho$ parameter of RooNDKeysPDF from 1 to 2; this increases the smoothing. The systematic uncertainty due to the background shape is studied by varying the relative fractions of the major backgrounds from $e^+e^- \to q\bar{q}$ and the non-$D_s^{*+}D^{-}_s$ open charm processes by $\pm30\%$, based on the errors  of input cross sections in the inclusive MC sample. Quadratic sum of the change of the fitted signal yield for each item, 1.6\%, is assigned as the systematic uncertainty in the $U_{\rm miss}$ fit.

\item  Least $|\Delta E|$. To estimate the systematic uncertainty in  the least $|\Delta E|$ method in the selection of the transition $\gamma(\pi^0)$ from $D^{*+}_s$, we use  the control samples of $D_s^+\to K^+K^-\pi^+$ and $D_s^+\to \eta\pi^0\pi^+$. The difference in the efficiencies of  selecting the transition  $\gamma(\pi^0)$ candidates  between data and MC simulation, 0.4\%, is assigned as the uncertainty. 

\item MC statistics.  The uncertainty due to MC statistics, 0.2\%, is assigned as a systematic uncertainty. 

\item MC model.  The systematic uncertainty of MC model is estimated by comparing the signal efficiency obtained with the  alternative signal MC samples generated  by varying the input FF ratios by $\pm 1 \sigma$ statistical error. The larger change of the signal efficiency, 0.8\%, is taken as the corresponding uncertainty.

\item  Quoted BF. The uncertainty of  the quoted BF of  $\phi\to K^{+}K^{-}$ is  1.0\%~\cite{ParticleDataGroup:2020ssz}. 
We have also examined the averaged signal efficiency  by varying the quoted BFs of $D^{*+}_s\to \gamma D^+_s$ and 
$D^{*+}_s\to \pi^0 D^+_s$  within $\pm 1\sigma$ and find the change of the signal efficiency is less than 0.2\%.  Adding these two items in quadrature gives the  total systematic uncertainty  due to the quoted BFs to be 1.0\%. 

The above sources are summarized in Table~\ref{tab:bf-syst-sum}. The total systematic uncertainty, obtained by summing the contributions in quadrature, is 3.3\%. 
\end{itemize}

\begin{table}[htp]
\centering
\caption{Relative systematic uncertainties  in the BF measurement. 
}
\begin{tabular}{lcc|c}
  \hline
  \hline
  Source                              &Uncertainty (\%)\\
  \hline
  $N^{\rm tot}_{\rm ST}$              & 0.5       \\
  Tracking of $K^\pm$                    & 1.8       \\
  PID of $K^\pm$                        & 0.8       \\
  Tracking of $\mu^+$                   & 0.3       \\
  PID of $\mu^+$                          & 0.4       \\
  Selection of transition  $\gamma(\pi^0)$       & 1.0 \\
  Tag bias                            & 0.1       \\
  $E^{\rm max}_{\rm extra \gamma}$, $N_{\rm extra}^{\rm char}$ and $N_{\rm extra}^{\pi^0}$                           &0.4        \\
  $M_{\phi \nu_{\mu}}$ and $M_{\phi \mu^+}$ & 1.0 \\
  ${ U_{\rm miss}}$ fit               & 1.6       \\
  Least $|\Delta E|$               & 0.4       \\
  MC statistics                       & 0.2       \\
  MC model                            & 0.8       \\
  Quoted BF                           & 1.0       \\
  \hline
  Total                               & 3.3       \\

  \hline
  \hline
\end{tabular}
\label{tab:bf-syst-sum}
\end{table}
\subsection{Measurement of form factors }
The following sources of systematic uncertainties, summarized in Table~\ref{tab:ff-syst-sum}, have been considered in the FF ratio measurements.
\begin{itemize}

\item 
Background estimation. 
First, the fractions of backgrounds for the two sample groups, i.e.~$f_{b}$ 
in Eq.~(\ref{eq:nll}), are varied by their corresponding statistical 
uncertainties, addressing background levels.  
Second, an alternative MC-simulated shape is used to examine the uncertainty 
arising from the background shape modeling. 
Alternative background shapes are obtained with the relative fractions 
of the backgrounds from $e^+e^- \to q\bar{q}$ and non-$D_s^{*+}D^{-}_s$ 
varied by the statistical uncertainties of their cross sections.  
The differences caused by these variations are assigned as the uncertainties. 

\item
$r_{BW}$.  The effective radius of the resonance is set to 3.0 $({\rm GeV}/c)^{-1}$ in the nominal fit. This  value is  varied from 1.0 $({\rm GeV}/c)^{-1}$ to 5.0 $({\rm GeV}/c)^{-1}$, taking the largest difference in the results as the systematic uncertainties. 

\item
$m_V$ and $m_A$. The fixed parameters $m_V$ and $m_A$ are varied by $\pm 100 ~ \text{MeV}/c^2$ to estimate the uncertainties associated with the pole mass assumption. 
The differences from the nominal result are assigned as systematic uncertainties.  

\item
$\phi$ line shape. The uncertainty is estimated by varying the mass and  width of the $\phi$ meson by $\pm1\sigma$; the largest difference is taken as systematic uncertainty. 
 
\item
Efficiency corrections.  These corrections compensate for efficiency differences between data and MC simulation from PID and tracking, reflected in the $\gamma_\epsilon$ parameters in Eq.~(\ref{neweq:sigmc-integral}). The uncertainties due to the $\gamma_\epsilon$ parameters are obtained by performing the PWA while varying PID and tracking efficiencies by their uncertainties.  The difference from the nominal result is assigned as systematic uncertainty. 

\end{itemize}

\begin{table}[htp]
\begin{center}
\caption{ Relative systematic uncertainties  of the measurements of  the FF ratios. }
\vspace{0.25cm}
\begin{tabular}{l|c|c}
\hline
\hline
Source                                 & $r_V$  & $r_2$  \\
\hline

Background  estimation                 & 0.31\% & 0.49\% \\
$r_{BW}$                               & 0.06\% & 0.28\% \\
$m_V$                                  & 0.95\% & 0.03\% \\
$m_A$                                  & 1.10\% & 2.39\% \\
$\phi$ line shape                      & 0.01\% & 0.07\% \\
Efficiency corrections                 & 0.13\% & 0.28\% \\
\hline
Total                                  & 1.46\% & 2.47\% \\
\hline
\hline
\end{tabular}
\label{tab:ff-syst-sum}
\end{center}
\end{table}

\section{Summary}
A PWA is performed on the SL decay $\kkmv$ for the first time using $7.33~\mathrm{fb}^{-1}$ of $e^+e^-$ collision data collected by the BESIII detector at $E_{\rm CM}$  in the range from  $4.128~$GeV to $4.226~$GeV.  
The absolute BF of $\phimv$ is  measured as $(2.25\pm 0.09 \pm 0.07) \times10^{-2}$. The precision of the BF is a factor of 4.3 better than the world average value.
Combining this result with the world average of ${\mathcal B}(D^+_s\to \phi e^+\nu_e)$~\cite{ParticleDataGroup:2020ssz}, the ratio of the two BFs obtained is  ${\mathcal B}(D^+_s\to\phi \mu^+\nu_\mu)/{\mathcal B}(D^+_s\to\phi e^+\nu_e) = 0.94 \pm 0.08$, consistent with the SM prediction.  
Assuming that the only $S$-wave contribution is from the $f_0(980)$, the process of $D_s^+\to f_0(980)\mu^{+}{\nu}_{\mu}, f_0(980)\to K^+K^-$ was searched for and no significant signal was found.  The upper limit $\mathcal{B}(D_s^+\to f_0(980)\mu^{+}{\nu}_{\mu}) \cdot {\mathcal B}( f_0(980)\to K^+K^-) < 5.45 \times 10^{-4}$ is set at the 90\% confidence level.

By assuming only the $\phi$ contribution, the FF ratios $r_{V}=\frac{V(0)}{A_{1}(0)}=1.58\pm0.17\pm0.02$ and $r_{2}=\frac{A_{2}(0)}{A_{1}(0)}=0.71\pm0.14\pm0.02$ are extracted.
These FFs measurements are summarized in Table~\ref{tab:com} and compared to the previous measurements and the theoretical calculations.

\begin{table}[htp]
\begin{center}
\caption{  Measured FF ratios and comparison with previous measurements.
}
\scalebox{0.80}{
\vspace{0.25cm}
\begin{tabular}{lccc}
\hline
Experiments                        & $r_V$                       & $r_2$                       \\
\hline
PDG~\cite{ParticleDataGroup:2020ssz}              &  1.80$\pm$0.08             & 0.84$\pm$0.11           \\
\hline
This analysis                      &  1.58$\pm$0.17$\pm$0.02     & 0.71$\pm$0.14$\pm$0.02     \\
\hline
$BABAR$~\cite{BaBar:2008gpr}        &  1.807$\pm$0.046$\pm$0.065  & 0.816$\pm$0.036$\pm$0.030  \\
\hline                                         
FOCUS~\cite{FOCUS:2004gfa}       &  1.549$\pm$0.250$\pm$0.148  & 0.713$\pm$0.202$\pm$0.284 \\                         
\hline                               
Theory                     & $r_V$                       & $r_2$            \\ 
\hline                                             
CCQM~\cite{Soni:2018adu}              & 1.34$\pm0.27$                   & 0.99$\pm$0.20                 \\
\hline                                             
CQM~\cite{Melikhov:2000yu}              &  1.72                   & 0.73                \\
\hline                                             
LFQM~\cite{Verma:2011yw}         & 1.42                            & 0.86                          \\
\hline                               
LQCD~\cite{Donald:2013pea}                   & 1.72$\pm0.21$                   & 0.74$\pm0.12$                 \\
\hline                                             
HM$\chi$T~\cite{Fajfer:2005ug}                    & 1.80                            & 0.52                        \\
\hline
\end{tabular}
}
\label{tab:com}
\end{center}
\end{table}

These FF measurements are consistent with the $BABAR$~\cite{BaBar:2008gpr} and FOCUS~\cite{FOCUS:2004gfa}  measurements. 
The obtained  FF ratios confirm the theoretical predictions~\cite{Soni:2018adu,Melikhov:2000yu,Verma:2011yw,Donald:2013pea}, which have been used in the determination of $|V_{cs}|$ and CKM unitarity tests.

\acknowledgments

The BESIII Collaboration thanks the staff of BEPCII and the IHEP computing center for their strong support. The authors thank Prof. Yao Yu for helpful discussions. 
This work is  supported in part by National Key R\&D Program of China under Contracts Nos. 2020YFA0406400, 2020YFA0406300; National Natural Science Foundation of China (NSFC) under Contracts Nos. 11805037, 11635010, 11735014, 11835012, 11935015, 11935016, 11935018, 11961141012, 12022510, 12025502, 12035009, 12035013, 12061131003, 12192260, 12192261, 12192262, 12192263, 12192264, 12192265, 12221005, 12225509, 12235017; Guangdong Basic and Applied Basic Research Foundation under Grant  No. 2023A1515010121; the Chinese Academy of Sciences (CAS) Large-Scale Scientific Facility Program; Joint Large-Scale Scientific Facility Funds of the NSFC and CAS under Contracts No. U2032104; the CAS Center for Excellence in Particle Physics (CCEPP); CAS Key Research Program of Frontier Sciences under Contracts Nos. QYZDJ-SSW-SLH003, QYZDJ-SSW-SLH040; 100 Talents Program of CAS; The Institute of Nuclear and Particle Physics (INPAC) and Shanghai Key Laboratory for Particle Physics and Cosmology;  ERC under Contract No. 758462; European Union's Horizon 2020 research and innovation programme under Marie Sklodowska-Curie grant agreement under Contract No. 894790; German Research Foundation DFG under Contracts Nos. 443159800, 455635585, Collaborative Research Center CRC 1044, FOR5327, GRK 2149; Istituto Nazionale di Fisica Nucleare, Italy; Ministry of Development of Turkey under Contract No. DPT2006K-120470; National Research Foundation of Korea under Contract No. NRF-2022R1A2C1092335; National Science and Technology fund of Mongolia; National Science Research and Innovation Fund (NSRF) via the Program Management Unit for Human Resources \& Institutional Development, Research and Innovation of Thailand under Contract No. B16F640076; Polish National Science Centre under Contract No. 2019/35/O/ST2/02907; The Swedish Research Council; U. S. Department of Energy under Contract No. DE-FG02-05ER41374.

\bibliographystyle{JHEP}
\bibliography{references}

\providecommand{\href}[2]{#2}\begingroup\raggedright\begin{thebibliography}{10}

\bibitem{Na:2011mc}
H.~Na, C.~T.~H. Davies, E.~Follana, J.~Koponen, G.~P. Lepage and J.~Shigemitsu,
  \emph{{$D \rightarrow \pi, l \nu$ Semileptonic Decays, $|V_{cd}|$ and
  2$^{nd}$ Row Unitarity from Lattice QCD}},
  \href{https://doi.org/10.1103/PhysRevD.84.114505}{\emph{Phys. Rev. D}
  {\bfseries 84} (2011) 114505}
  [\href{https://arxiv.org/abs/1109.1501}{{\ttfamily arXiv:1109.1501}}].

\bibitem{Aoki:2016frl}
S.~Aoki et~al., \emph{{Review of lattice results concerning low-energy particle
  physics}}, \href{https://doi.org/10.1140/epjc/s10052-016-4509-7}{\emph{Eur.
  Phys. J. C} {\bfseries 77} (2017) 112}
  [\href{https://arxiv.org/abs/1607.00299}{{\ttfamily arXiv:1607.00299}}].

\bibitem{Donald:2013pea}
{\scshape HPQCD} collaboration, \emph{{$V_{cs}$ from $D_s \to \phi \ell \nu$
  semileptonic decay and full lattice QCD}},
  \href{https://doi.org/10.1103/PhysRevD.90.074506}{\emph{Phys. Rev. D}
  {\bfseries 90} (2014) 074506}
  [\href{https://arxiv.org/abs/1311.6669}{{\ttfamily arXiv:1311.6669}}].

\bibitem{Donald:2013kla}
{\scshape HPQCD} collaboration, \emph{{$D_s$ to $\phi$and other transitions
  from lattice QCD}}, \href{https://doi.org/10.22323/1.187.0390}{\emph{PoS}
  {\bfseries LATTICE2013} (2014) 390}
  [\href{https://arxiv.org/abs/1311.6967}{{\ttfamily arXiv:1311.6967}}].

\bibitem{Soni:2018adu}
N.~R. Soni, M.~A. Ivanov, J.~G. K\"orner, J.~N. Pandya, P.~Santorelli and C.~T.
  Tran, \emph{{Semileptonic $D_{(s)}$-meson decays in the light of recent
  data}}, \href{https://doi.org/10.1103/PhysRevD.98.114031}{\emph{Phys. Rev. D}
  {\bfseries 98} (2018) 114031}
  [\href{https://arxiv.org/abs/1810.11907}{{\ttfamily arXiv:1810.11907}}].

\bibitem{Melikhov:2000yu}
D.~Melikhov and B.~Stech, \emph{{Weak form-factors for heavy meson decays: An
  Update}}, \href{https://doi.org/10.1103/PhysRevD.62.014006}{\emph{Phys. Rev.
  D} {\bfseries 62} (2000) 014006}
  [\href{https://arxiv.org/abs/hep-ph/0001113}{{\ttfamily hep-ph/0001113}}].

\bibitem{Verma:2011yw}
R.~C. Verma, \emph{{Decay constants and form factors of s-wave and p-wave
  mesons in the covariant light-front quark model}},
  \href{https://doi.org/10.1088/0954-3899/39/2/025005}{\emph{J. Phys. G}
  {\bfseries 39} (2012) 025005}
  [\href{https://arxiv.org/abs/1103.2973}{{\ttfamily arXiv:1103.2973}}].

\bibitem{Fajfer:2005ug}
S.~Fajfer and J.~F. Kamenik, \emph{{Charm meson resonances and D $\to$ V
  semileptonic form-factors}},
  \href{https://doi.org/10.1103/PhysRevD.72.034029}{\emph{Phys. Rev. D}
  {\bfseries 72} (2005) 034029}
  [\href{https://arxiv.org/abs/hep-ph/0506051}{{\ttfamily hep-ph/0506051}}].

\bibitem{LHCb:2021trn}
{\scshape LHCb} collaboration, \emph{{Test of lepton universality in
  beauty-quark decays}},
  \href{https://doi.org/10.1038/s41567-021-01478-8}{\emph{Nature Phys.}
  {\bfseries 18} (2022) 277}
  [\href{https://arxiv.org/abs/2103.11769}{{\ttfamily arXiv:2103.11769}}].

\bibitem{LHCb:2022zom}
{\scshape LHCb} collaboration, \emph{{Measurement of lepton universality
  parameters in $B^+\to K^+\ell^+\ell^-$ and $B^0\to K^{*0}\ell^+\ell^-$
  decays}},  \href{https://arxiv.org/abs/2212.09153}{{\ttfamily
  arXiv:2212.09153}}.

\bibitem{Bauer:2015knc}
M.~Bauer and M.~Neubert, \emph{{Minimal Leptoquark Explanation for the
  $R_{D^{(*)}}$ , $R_K$ , and $(g-2)_\mu$ Anomalies}},
  \href{https://doi.org/10.1103/PhysRevLett.116.141802}{\emph{Phys. Rev. Lett.}
  {\bfseries 116} (2016) 141802}
  [\href{https://arxiv.org/abs/1511.01900}{{\ttfamily arXiv:1511.01900}}].

\bibitem{Crivellin:2015hha}
A.~Crivellin, J.~Heeck and P.~Stoffer, \emph{{A perturbed lepton-specific
  two-Higgs-doublet model facing experimental hints for physics beyond the
  Standard Model}},
  \href{https://doi.org/10.1103/PhysRevLett.116.081801}{\emph{Phys. Rev. Lett.}
  {\bfseries 116} (2016) 081801}
  [\href{https://arxiv.org/abs/1507.07567}{{\ttfamily arXiv:1507.07567}}].

\bibitem{Crivellin:2015mga}
A.~Crivellin, G.~D'Ambrosio and J.~Heeck, \emph{{Explaining
  $h\to\mu^\pm\tau^\mp$, $B\to K^* \mu^+\mu^-$ and $B\to K \mu^+\mu^-/B\to K
  e^+e^-$ in a two-Higgs-doublet model with gauged $L_\mu-L_\tau$}},
  \href{https://doi.org/10.1103/PhysRevLett.114.151801}{\emph{Phys. Rev. Lett.}
  {\bfseries 114} (2015) 151801}
  [\href{https://arxiv.org/abs/1501.00993}{{\ttfamily arXiv:1501.00993}}].

\bibitem{Fajfer:2012jt}
S.~Fajfer, J.~F. Kamenik, I.~Nisandzic and J.~Zupan, \emph{{Implications of
  Lepton Flavor Universality Violations in B Decays}},
  \href{https://doi.org/10.1103/PhysRevLett.109.161801}{\emph{Phys. Rev. Lett.}
  {\bfseries 109} (2012) 161801}
  [\href{https://arxiv.org/abs/1206.1872}{{\ttfamily arXiv:1206.1872}}].

\bibitem{Fajfer:2012vx}
S.~Fajfer, J.~F. Kamenik and I.~Nisandzic, \emph{{On the $B \to D^* \tau \bar
  \nu_{\tau}$ Sensitivity to New Physics}},
  \href{https://doi.org/10.1103/PhysRevD.85.094025}{\emph{Phys. Rev. D}
  {\bfseries 85} (2012) 094025}
  [\href{https://arxiv.org/abs/1203.2654}{{\ttfamily arXiv:1203.2654}}].

\bibitem{BESIII:2018nzb}
{\scshape BESIII} collaboration, \emph{{Measurement of the branching fraction
  for the semi-leptonic decay $D^{0(+)}\to \pi^{-(0)}\mu^+\nu_\mu$ and test of
  lepton universality}},
  \href{https://doi.org/10.1103/PhysRevLett.121.171803}{\emph{Phys. Rev. Lett.}
  {\bfseries 121} (2018) 171803}
  [\href{https://arxiv.org/abs/1802.05492}{{\ttfamily arXiv:1802.05492}}].

\bibitem{BESIII:2019gsm}
{\scshape BESIII} collaboration, \emph{{Study of the $D^0 \rightarrow K^- \mu^+
  \nu_\mu$ Dynamics and Test of Lepton Flavor Universality with $D^0
  \rightarrow K^- \ell^+ \nu_\ell$ Decays}},
  \href{https://doi.org/10.1142/9789811217739_0041}{\emph{Phys. Rev. Lett.}
  {\bfseries 122} (2019) 011804}
  [\href{https://arxiv.org/abs/1810.03127}{{\ttfamily arXiv:1810.03127}}].

\bibitem{Ablikim:2020hsc}
{\scshape BESIII} collaboration, \emph{{First Observation of $D^+ \rightarrow
  \eta\mu^+\nu_\mu$ and Measurement of Its Decay Dynamics}},
  \href{https://doi.org/10.1103/PhysRevLett.124.231801}{\emph{Phys. Rev. Lett.}
  {\bfseries 124} (2020) 231801}
  [\href{https://arxiv.org/abs/2003.12220}{{\ttfamily arXiv:2003.12220}}].

\bibitem{BESIII:2019qci}
{\scshape BESIII} collaboration, \emph{{Measurement of the Dynamics of the
  Decays $D_s^+ \rightarrow \eta^{(\prime)} e^+ \nu_e$}},
  \href{https://doi.org/10.1103/PhysRevLett.122.121801}{\emph{Phys. Rev. Lett.}
  {\bfseries 122} (2019) 121801}
  [\href{https://arxiv.org/abs/1901.02133}{{\ttfamily arXiv:1901.02133}}].

\bibitem{BESIII:2017ikf}
{\scshape BESIII} collaboration, \emph{{Measurements of the branching fractions
  for the semi-leptonic decays $D^+_s\to\phi e^{+}\nu_{e}$, $\phi
  \mu^{+}\nu_{\mu}$, $\eta \mu^{+}\nu_{\mu}$ and $\eta'\mu^{+}\nu_{\mu}$}},
  \href{https://doi.org/10.1103/PhysRevD.97.012006}{\emph{Phys. Rev. D}
  {\bfseries 97} (2018) 012006}
  [\href{https://arxiv.org/abs/1709.03680}{{\ttfamily arXiv:1709.03680}}].

\bibitem{BESIII:2021ynj}
{\scshape BESIII} collaboration, \emph{{First Measurement of the Absolute
  Branching Fraction of $\Lambda \to p \mu^- \bar{\nu}_{\mu}$}},
  \href{https://doi.org/10.1103/PhysRevLett.127.121802}{\emph{Phys. Rev. Lett.}
  {\bfseries 127} (2021) 121802}
  [\href{https://arxiv.org/abs/2107.06704}{{\ttfamily arXiv:2107.06704}}].

\bibitem{Belle:2021crz}
{\scshape Belle} collaboration, \emph{{Measurements of the branching fractions
  of the semileptonic decays $\Xi_{c}^{0} \to \Xi^{-} \ell^{+} \nu_{\ell}$ and
  the asymmetry parameter of $\Xi_{c}^{0} \to \Xi^{-} \pi^{+}$}},
  \href{https://doi.org/10.1103/PhysRevLett.127.121803}{\emph{Phys. Rev. Lett.}
  {\bfseries 127} (2021) 121803}
  [\href{https://arxiv.org/abs/2103.06496}{{\ttfamily arXiv:2103.06496}}].

\bibitem{Belle:2021dgc}
{\scshape Belle} collaboration, \emph{{First test of lepton flavor universality
  in the charmed baryon decays $\Omega_{c}^{0} \to \Omega^{-} \ell^{+}
  \nu_{\ell}$ using data of the Belle experiment}},
  \href{https://doi.org/10.1103/PhysRevD.105.L091101}{\emph{Phys. Rev. D}
  {\bfseries 105} (2022) L091101}
  [\href{https://arxiv.org/abs/2112.10367}{{\ttfamily arXiv:2112.10367}}].

\bibitem{BESIII:2015ysy}
{\scshape BESIII} collaboration, \emph{{Measurement of the absolute branching
  fraction for $\Lambda^+_{c}\to \Lambda e^+\nu_e$}},
  \href{https://doi.org/10.1103/PhysRevLett.115.221805}{\emph{Phys. Rev. Lett.}
  {\bfseries 115} (2015) 221805}
  [\href{https://arxiv.org/abs/1510.02610}{{\ttfamily arXiv:1510.02610}}].

\bibitem{BaBar:2008gpr}
{\scshape BaBar} collaboration, \emph{{Study of the decay $D^+_{s} \to K^{+}
  K^{-} e^{+} \nu_{e}$}},
  \href{https://doi.org/10.1103/PhysRevD.78.051101}{\emph{Phys. Rev. D}
  {\bfseries 78} (2008) 051101}
  [\href{https://arxiv.org/abs/0807.1599}{{\ttfamily arXiv:0807.1599}}].

\bibitem{Kim:2018zob}
H.~Kim, K.~S. Kim, M.-K. Cheoun, D.~Jido and M.~Oka, \emph{{Further signatures
  to support the tetraquark mixing framework for the two light-meson nonets}},
  \href{https://doi.org/10.1103/PhysRevD.99.014005}{\emph{Phys. Rev. D}
  {\bfseries 99} (2019) 014005}
  [\href{https://arxiv.org/abs/1811.00187}{{\ttfamily arXiv:1811.00187}}].

\bibitem{Wang:2022vga}
Z.-Q. Wang, X.-W. Kang, J.~A. Oller and L.~Zhang, \emph{{Analysis on the
  composite nature of the light scalar mesons $f_0(980)$ and $a_0(980)$}},
  \href{https://doi.org/10.1103/PhysRevD.105.074016}{\emph{Phys. Rev. D}
  {\bfseries 105} (2022) 074016}
  [\href{https://arxiv.org/abs/2201.00492}{{\ttfamily arXiv:2201.00492}}].

\bibitem{Stone:2013eaa}
S.~Stone and L.~Zhang, \emph{{Use of $B\to J/\psi f_0$ decays to discern the $q
  \bar{q}$ or tetraquark nature of scalar mesons}},
  \href{https://doi.org/10.1103/PhysRevLett.111.062001}{\emph{Phys. Rev. Lett.}
  {\bfseries 111} (2013) 062001}
  [\href{https://arxiv.org/abs/1305.6554}{{\ttfamily arXiv:1305.6554}}].

\bibitem{Yao:2020bxx}
D.-L. Yao, L.-Y. Dai, H.-Q. Zheng and Z.-Y. Zhou, \emph{{A review on
  partial-wave dynamics with chiral effective field theory and dispersion
  relation}}, \href{https://doi.org/10.1088/1361-6633/abfa6f}{\emph{Rept. Prog.
  Phys.} {\bfseries 84} (2021) 076201}
  [\href{https://arxiv.org/abs/2009.13495}{{\ttfamily arXiv:2009.13495}}].

\bibitem{Dai:2014zta}
L.-Y. Dai and M.~R. Pennington, \emph{{Comprehensive amplitude analysis of
  $\gamma\gamma \rightarrow \pi^+\pi^-, \pi^0\pi^0$ and $\overline{K} K$ below
  1.5 GeV}}, \href{https://doi.org/10.1103/PhysRevD.90.036004}{\emph{Phys. Rev.
  D} {\bfseries 90} (2014) 036004}
  [\href{https://arxiv.org/abs/1404.7524}{{\ttfamily arXiv:1404.7524}}].

\bibitem{BESIII:2009fln}
{\scshape BESIII} collaboration, \emph{{Design and Construction of the BESIII
  Detector}}, \href{https://doi.org/10.1016/j.nima.2009.12.050}{\emph{Nucl.
  Instrum. Meth. A} {\bfseries 614} (2010) 345}
  [\href{https://arxiv.org/abs/0911.4960}{{\ttfamily arXiv:0911.4960}}].

\bibitem{Yu:2016cof}
C.~Yu et~al., \emph{{BEPCII Performance and Beam Dynamics Studies on
  Luminosity}},  in \emph{{7th International Particle Accelerator Conference}},
  p.~TUYA01, 2016, \href{https://doi.org/10.18429/JACoW-IPAC2016-TUYA01}{DOI}.

\bibitem{BESIII:2020nme}
{\scshape BESIII} collaboration, \emph{{Future Physics Programme of BESIII}},
  \href{https://doi.org/10.1088/1674-1137/44/4/040001}{\emph{Chin. Phys. C}
  {\bfseries 44} (2020) 040001}
  [\href{https://arxiv.org/abs/1912.05983}{{\ttfamily arXiv:1912.05983}}].

\bibitem{Li:2021iwf}
H.-B. Li and X.-R. Lyu, \emph{{Study of the standard model with weak decays of
  charmed hadrons at BESIII}},
  \href{https://doi.org/10.1093/nsr/nwab181}{\emph{Natl. Sci. Rev.} {\bfseries
  8} (2021) nwab181} [\href{https://arxiv.org/abs/2103.00908}{{\ttfamily
  arXiv:2103.00908}}].

\bibitem{Huang:2022wuo}
K.-X. Huang, Z.-J. Li, Z.~Qian, J.~Zhu, H.-Y. Li, Y.-M. Zhang et~al.,
  \emph{{Method for detector description transformation to Unity and
  application in BESIII}},
  \href{https://doi.org/10.1007/s41365-022-01133-8}{\emph{Nucl. Sci. Tech.}
  {\bfseries 33} (2022) 142}
  [\href{https://arxiv.org/abs/2206.10117}{{\ttfamily arXiv:2206.10117}}].

\bibitem{Cao:2020ibk}
P.~Cao et~al., \emph{{Design and construction of the new BESIII endcap
  Time-of-Flight system with MRPC Technology}},
  \href{https://doi.org/10.1016/j.nima.2019.163053}{\emph{Nucl. Instrum. Meth.
  A} {\bfseries 953} (2020) 163053}.

\bibitem{GEANT4:2002zbu}
{\scshape GEANT4} collaboration, \emph{{GEANT4--a simulation toolkit}},
  \href{https://doi.org/10.1016/S0168-9002(03)01368-8}{\emph{Nucl. Instrum.
  Meth. A} {\bfseries 506} (2003) 250}.

\bibitem{Jadach:2000ir}
S.~Jadach, B.~F.~L. Ward and Z.~Was, \emph{{Coherent exclusive exponentiation
  for precision Monte Carlo calculations}},
  \href{https://doi.org/10.1103/PhysRevD.63.113009}{\emph{Phys. Rev. D}
  {\bfseries 63} (2001) 113009}
  [\href{https://arxiv.org/abs/hep-ph/0006359}{{\ttfamily hep-ph/0006359}}].

\bibitem{Ping:2013jka}
R.-G. Ping, \emph{{An exclusive event generator for $e^+ e^-$ scan
  experiments}},
  \href{https://doi.org/10.1088/1674-1137/38/8/083001}{\emph{Chin. Phys. C}
  {\bfseries 38} (2014) 083001}
  [\href{https://arxiv.org/abs/1309.3932}{{\ttfamily arXiv:1309.3932}}].

\bibitem{Lange:2001uf}
D.~J. Lange, \emph{{The EvtGen particle decay simulation package}},
  \href{https://doi.org/10.1016/S0168-9002(01)00089-4}{\emph{Nucl. Instrum.
  Meth. A} {\bfseries 462} (2001) 152}.

\bibitem{Ping:2008zz}
R.-G. Ping, \emph{{Event generators at BESIII}},
  \href{https://doi.org/10.1088/1674-1137/32/8/001}{\emph{Chin. Phys. C}
  {\bfseries 32} (2008) 599}.

\bibitem{ParticleDataGroup:2020ssz}
{\scshape Particle Data Group} collaboration, \emph{{Review of Particle
  Physics}}, \href{https://doi.org/10.1093/ptep/ptaa104}{\emph{PTEP} {\bfseries
  2022} (2022) 083C01}.

\bibitem{Chen:2000tv}
J.~C. Chen, G.~S. Huang, X.~R. Qi, D.~H. Zhang and Y.~S. Zhu, \emph{{Event
  generator for J / psi and psi (2S) decay}},
  \href{https://doi.org/10.1103/PhysRevD.62.034003}{\emph{Phys. Rev. D}
  {\bfseries 62} (2000) 034003}.

\bibitem{Yang:2014vra}
R.-L. Yang, R.-G. Ping and H.~Chen, \emph{{Tuning and Validation of the
  Lundcharm Model with $J/\psi$ Decays}},
  \href{https://doi.org/10.1088/0256-307X/31/6/061301}{\emph{Chin. Phys. Lett.}
  {\bfseries 31} (2014) 061301}.

\bibitem{Richter-Was:1992hxq}
E.~Richter-Was, \emph{{QED bremsstrahlung in semileptonic B and leptonic tau
  decays}}, \href{https://doi.org/10.1016/0370-2693(93)90062-M}{\emph{Phys.
  Lett. B} {\bfseries 303} (1993) 163}.

\bibitem{BESIII:2021anh}
{\scshape BESIII} collaboration, \emph{{Measurement of the absolute branching
  fractions for purely leptonic $D_s^+$ decays}},
  \href{https://doi.org/10.1103/PhysRevD.104.052009}{\emph{Phys. Rev. D}
  {\bfseries 104} (2021) 052009}
  [\href{https://arxiv.org/abs/2102.11734}{{\ttfamily arXiv:2102.11734}}].

\bibitem{MARK-III:1985hbd}
{\scshape MARK-III} collaboration, \emph{{Direct Measurements of Charmed d
  Meson Hadronic Branching Fractions}},
  \href{https://doi.org/10.1103/PhysRevLett.56.2140}{\emph{Phys. Rev. Lett.}
  {\bfseries 56} (1986) 2140}.

\bibitem{BESIII:2018hhz}
{\scshape BESIII} collaboration, \emph{{Determination of the pseudoscalar decay
  constant $f_{D_s^+}$ via $D_s^+\to\mu^+\nu_\mu$}},
  \href{https://doi.org/10.1103/PhysRevLett.122.071802}{\emph{Phys. Rev. Lett.}
  {\bfseries 122} (2019) 071802}
  [\href{https://arxiv.org/abs/1811.10890}{{\ttfamily arXiv:1811.10890}}].

\bibitem{BESIII:2019pjk}
{\scshape BESIII} collaboration, \emph{{Search for the decay $D_s^+\rightarrow
  \gamma e^+\nu_e$}},
  \href{https://doi.org/10.1103/PhysRevD.99.072002}{\emph{Phys. Rev. D}
  {\bfseries 99} (2019) 072002}
  [\href{https://arxiv.org/abs/1902.03351}{{\ttfamily arXiv:1902.03351}}].

\bibitem{Ablikim:2009aa}
M.~Ablikim, Z.~An, J.~Bai, N.~Berger, J.~Bian, X.~Cai et~al., \emph{Design and
  construction of the {BESIII} detector},
  \href{https://doi.org/https://doi.org/10.1016/j.nima.2009.12.050}{\emph{Nucl.
  Instrum. Meth. A} {\bfseries 614} (2010) 345}.

\bibitem{Cabibbo:1965zzb}
N.~Cabibbo and A.~Maksymowicz, \emph{{Angular Correlations in $K_{e4}$ Decays
  and Determination of Low-Energy $\pi$ - $\pi$ Phase Shifts}},
  \href{https://doi.org/10.1103/PhysRev.137.B438}{\emph{Phys. Rev.} {\bfseries
  137} (1965) B438} [Erratum: Phys.Rev. 168, 1926 (1968)].

\bibitem{Lee:1992ih}
C.~L.~Y. Lee, M.~Lu and M.~B. Wise, \emph{{B(l4) and D(l4) decay}},
  \href{https://doi.org/10.1103/PhysRevD.46.5040}{\emph{Phys. Rev. D}
  {\bfseries 46} (1992) 5040}.

\bibitem{Zhang:2023nnn}
H.~Zhang, B.-C. Ke, Y.~Yu and E.~Wang, \emph{{Lepton mass correction in partial
  wave analyses of charmed meson semi-leptonic decays*}},
  \href{https://doi.org/10.1088/1674-1137/acc642}{\emph{Chin. Phys. C}
  {\bfseries 47} (2023) 063101}
  [\href{https://arxiv.org/abs/2302.10541}{{\ttfamily arXiv:2302.10541}}].

\bibitem{BES:2004twe}
{\scshape BES} collaboration, \emph{{Resonances in $J/\psi \to \phi \pi^{+}
  \pi^{-}$ and $\phi K^{+} K^{-}$}},
  \href{https://doi.org/10.1016/j.physletb.2004.12.041}{\emph{Phys. Lett. B}
  {\bfseries 607} (2005) 243}
  [\href{https://arxiv.org/abs/hep-ex/0411001}{{\ttfamily hep-ex/0411001}}].

\bibitem{CLEO:2012beo}
{\scshape CLEO} collaboration, \emph{{Amplitude analysis of $D^0\to
  K^+K^-\pi^+\pi^-$}},
  \href{https://doi.org/10.1103/PhysRevD.85.122002}{\emph{Phys. Rev. D}
  {\bfseries 85} (2012) 122002}
  [\href{https://arxiv.org/abs/1201.5716}{{\ttfamily arXiv:1201.5716}}].

\bibitem{Cranmer:2000du}
K.~S. Cranmer, \emph{{Kernel estimation in high-energy physics}},
  \href{https://doi.org/10.1016/S0010-4655(00)00243-5}{\emph{Comput. Phys.
  Commun.} {\bfseries 136} (2001) 198}
  [\href{https://arxiv.org/abs/hep-ex/0011057}{{\ttfamily hep-ex/0011057}}].

\bibitem{FOCUS:2004gfa}
{\scshape FOCUS} collaboration, \emph{{New measurements of the $D^+_{s} \to
  \phi \mu^{+} \nu_{\mu}$ form-factor ratios}},
  \href{https://doi.org/10.1016/j.physletb.2004.02.015}{\emph{Phys. Lett. B}
  {\bfseries 586} (2004) 183}
  [\href{https://arxiv.org/abs/hep-ex/0401001}{{\ttfamily hep-ex/0401001}}].

\end{thebibliography}\endgroup
\clearpage
\appendix
%% Saved at => 2022-12-07
\large
The BESIII Collaboration\\
\normalsize
\\M.~Ablikim$^{1}$, M.~N.~Achasov$^{13,b}$, P.~Adlarson$^{75}$, X.~C.~Ai$^{81}$, R.~Aliberti$^{36}$, A.~Amoroso$^{74A,74C}$, M.~R.~An$^{40}$, Q.~An$^{71,58}$, Y.~Bai$^{57}$, O.~Bakina$^{37}$
% ...
\\
\vspace{0.2cm}
%(BESIII Collaboration)\\
\vspace{0.2cm} {\it
$^{1}$ Institute of High Energy Physics, Beijing 100049, People's Republic of China\\
% ...
$^{13}$ G.I. Budker Institute of Nuclear Physics SB RAS (BINP), Novosibirsk 630090, Russia\\
% ...
$^{36}$ Johannes Gutenberg University of Mainz, Johann-Joachim-Becher-Weg 45, D-55099 Mainz, Germany\\
$^{37}$ Joint Institute for Nuclear Research, 141980 Dubna, Moscow region, Russia\\
% ...
$^{40}$ Liaoning Normal University, Dalian 116029, People's Republic of China\\
% ...
$^{57}$ Southeast University, Nanjing 211100, People's Republic of China\\
$^{58}$ State Key Laboratory of Particle Detection and Electronics, Beijing 100049, Hefei 230026, People's Republic of China\\
% ...
$^{71}$ University of Science and Technology of China, Hefei 230026, People's Republic of China\\
% ...
$^{74}$ University of Turin and INFN, (A)University of Turin, I-10125, Turin, Italy; (B)University of Eastern Piedmont, I-15121, Alessandria, Italy; (C)INFN, I-10125, Turin, Italy\\
$^{75}$ Uppsala University, Box 516, SE-75120 Uppsala, Sweden\\
% ...
$^{81}$ Zhengzhou University, Zhengzhou 450001, People's Republic of China\\
\vspace{0.2cm}
% ...
$^{b}$ Also at the Novosibirsk State University, Novosibirsk, 630090, Russia\\
% ...
}
%% ends here %%
\end{document}